%% file: mf.tex
\documentclass[12pt]{article}
\usepackage{graphicx}
\usepackage{caption}
\usepackage{float}
\usepackage{amsmath,amsfonts,amssymb}
\usepackage{xcolor}
\usepackage{dcolumn}% Align table columns on decimal point
\usepackage{tabularx}
\usepackage{longtable}
\usepackage{multirow}
\usepackage{relsize}
\usepackage[thinspace, thinqspace,amssymb]{SIunits}
\usepackage[version=4]{mhchem}
\usepackage{subfigure}
\usepackage{placeins}
\usepackage{acronym}
\usepackage[hidelinks]{hyperref}
\usepackage{lipsum}
\usepackage{pgfplots}
\usepackage{pgfplotstable}
\usepackage{tikz}
\pgfplotsset{/pgf/number format/use comma, compat=newest}
\usepgfplotslibrary{external,colormaps}
\tikzsetexternalprefix{figures/}

\usetikzlibrary{shapes,patterns,positioning,calc,decorations.pathreplacing,plotmarks}

\begin{document}

%\begin{frontmatter}



\title{{\sc Combined Notch and Size Effect Modeling in a Local Probabilistic Approach for LCF}}


\author{{\sc L. M\"ade, S. Schmitz, H. Gottschalk and T. Beck}}
\maketitle
\begin{abstract}
%A local probabilistic model for low cycle fatigue (LCF) based on the so-called size effect has been applied on gas-turbine design in recent years. It was also extended for combined notch support and size effect modeling and recently applied to a turbine vane. The notch support extension based on the stress gradient effect is described in detail here as well as the correct parameter calibration. Although the notch support itself requires just two additional parameters, modeling the combined effect requires simultaneous calibration of all model parameters. This was conducted for data of the superalloys \textit{IN-939} and \textit{Rene80} as well as for the steel \textit{26NiCrMoV14-5} using LCF test data of standard and notched specimens. The calibrated models are able to predict the probabilistic W\"ohler curves further notched specimen and cooling hole specimen test data that has not been used for calibration. The improved LCF life prediction by combined notch support and size effect modeling increases the confidence for applying the tool in component design as it is able to assess the stress gradient effect in arbitrary geometries.

In recent years a local probabilistic model for low cycle fatigue (LCF) based on the statistical size effect has been developed and applied on engineering components. 
%It was also extended for combined notch support and size effect modeling and recently applied to a gas-turbine vane. 
Here, the notch support extension based on the stress gradient effect is described in detail, as well as an FEA-based parameter calibration. An FEA is necessary to simulate non-homogeneous stress fields in non-smooth specimens which exhibit gradients and determine size effects. The hazard density approach and the surface integration over the FEA stress lead to geometry-independent model parameters. Three different materials (superalloys \textit{IN-939}, \textit{Rene80}, steel \textit{26NiCrMoV14-5}) and three different geometry types (smooth, notch, cooling hole specimens) are considered for a more comprehensive validation of the probabilistic LCF model and to demonstrate its wide application range. At the same time, a reduced testing effort is needed compared to deterministic model predictions with notch support.
\end{abstract}

\noindent{\textbf{Keywords:}
LCF, Probabilistic Modeling, Size Effect, Notch Support, Coffin-Manson-Basquin model

%\end{frontmatter}

%%%%%%%%%%%%%%%%%%%%%%%%%%%%%%%%%%%%%%%%%%%%%%%%%%%%%%%%%%%%%%%%%%%%%%
\input{Introduction}
%Link to former publications/work. 
%Here: Focus on Validation/predictiveness with three materials, three specimen types (standard, notch, coolinghole)
%(from calibration to validation)
%Simple model and big prediction effects for notch support compared to: BAM-work Fedelich, AG Turbo (Seibel, Engel)

%%%%%%%%%%%%%%%%%%%%%%%%%%%%%%%%%%%%%%%%%%%%%%%%%%%%%%%%%%%%%%%%%%%%%%
\input{Theory}
%Theory on materials:
%differences between steels  \& superalloys (microstructure, properties like creep resistance, temperature resistance/ductility)
%consult Dirk on steel properties
% Zusammenhang zw. A,k und Korngröße. Können NS Parameter möglicherweise anhand der Korngröße abgeschätzt werden?

%%%%%%%%%%%%%%%%%%%%%%%%%%%%%%%%%%%%%%%%%%%%%%%%%%%%%%%%%%%%%%%%%%%%%%
\input{Experiments}
%-M0044 (sensitivity to fit/calibration data sets, discuss prediction curve spreading for sharp notches) (discuss design issues for SU: no notches sharper than Kt=1.6)

%%%%%%%%%%%%%%%%%%%%%%%%%%%%%%%%%%%%%%%%%%%%%%%%%%%%%%%%%%%%%%%%%%%%%%
\input{Discussion}

%%%%%%%%%%%%%%%%%%%%%%%%%%%%%%%%%%%%%%%%%%%%%%%%%%%%%%%%%%%%%%%%%%%%%%
\section*{Acknowledgments} 
Part of this work has been supported by the German federal ministry of economic affairs BMWi via an AG Turbo grant. We thank Phillip Gravett, Georg Rollmann and Sachin Shinde from the gas turbine technology department and Dirk Kulawinski from the steam turbine technology department of Siemens AG for providing test data and guidance as well as stimulating discussions.
%%%%%%%%%%%%%%%%%%%%%%%%%%%%%%%%%%%%%%%%%%%%%%%%%%%%%%%%%%%%%%%%%%%%%%

\section*{Permission of use statement}
The content of this paper is copyrighted by Siemens AG and is licensed to the publisher for publication and distribution only. Any inquiries regarding permission to use the content of this
paper, in whole or in part, for any purpose must be addressed to Siemens AG directly.
~
\\

\noindent Self archiving is permitted by Siemens AG.
%%%%%%%%%%%%%%%%%%%%%%%%%%%%%%%%%%%%%%%%%%%%%%%%%%%%%%%%%%%%%%%%%%%%%%



%% References


%\bibliographystyle{elsarticle-num}



\vspace{2cm}

\noindent {\sc Lucas M\"ade}\\
Corresponding author: Gas Turbine Department of Materials\\ and Technology, Siemens AG, Huttenstra\ss e 12,\\ 10553, Berlin, Germany, email: lucas.maede@siemens.com\\
~
\\
\noindent {\sc Sebastian Schmitz}\\
 Gas Turbine Department of Materials and Technology, \\ Siemens AG, Huttenstra\ss e 12, 10553, Berlin, Germany,\\ email: schmitz.sebastian@siemens.com\\
~
\\
%\address{c Siemens AG, Mellinghoferstra\ss e 55, 45473 M\"ulheim an der Ruhr, Germany, email: georg.rollmann@siemens.com}
{\sc Hanno Gottschalk}\\
University of Wuppertal,School of Mathematics\\ and Science,  Gau\ss stra\ss e 20, \\ 42119, Wuppertal, Germany,\\ email: hanno.gottschalk@uni-wuppertal.de\\
~
\\
{\sc Tilman Beck}\\
Institute of Materials Science and Engineering,\\ Technische Universit\"at Kaiserslautern,\\ P.O. Box 3049, 67653, Kaiserslautern, Germany,\\ email: beck@mv.uni-kl.de
\end{document}

%% file: Introduction.tex
\section{Introduction}\label{sec:Int}
%For the analyses of the mechanical behavior of materials, specimen testing plays a crucial role in order to obtain material properties such as fatigue behavior. There are different kinds of specimen geometries and testing conditions to focus on different material properties which are used to derive mechanical integrity estimations for engineering components under certain loading conditions. 
%Entfernt auf Empfehlung von Prof. Beck

Traditionally, a combination of deterministic fatigue prediction considering the peak stress in a component only and safety factors accommodating for inherent statistical scatter, emerging from various uncertainties in the material and load conditions, is used for life assessment of engineering designs. 

However, not only mechanical integrity analyses of components have to deal with uncertainties in fatigue life. Also cyclic-fatigue experiments in lab conditions with gas turbine blade and rotor disc materials, such as \ce{Ni}-base superalloys and martensitic steels show a high scatter \cite{Harders_Roesler2007,Vormwald2007}. 
%It can often be decreased via increased testing efforts and understanding of material properties on the micro- and mesoscale. But of course, the resources for specimen testing and for obtaining information on material properties of engineering components in service are usually limited. %Entfernt auf Empfehlung von Prof. Gottschalk & Prof. Beck
This creates the challenging task to develop material fatigue models which provide the characteristics of failure probability. 

For life assessment of arbitrarily shaped engineering components at arbitrary loading conditions with respect to the uncertainties in material fatigue mechanisms, lifing models with a local probabilistic approach have advantages compared to deterministic approaches. The local probabilistic approach enables the transfer of mechanical properties and integrity evaluations from one geometry to another, thereby delivering geometry-independent material parameters which are a key for significantly reducing testing efforts. 

In this work, the focus is on low-cycle fatigue (LCF). The probabilistic LCF model of \cite{Schmitz_Seibel2013,ASME2013Paper,ASME2017Paper} is extended to account for the combined size and notch support effect. A comprehensive validation of this extended model is conducted with three different materials and three different specimen geometry classes.

%The ongoing strive for increased performance and efficiency of engineering components such as turbomachinery design goes often hand in hand with higher demands on the mechanical integrity of the components. 
The ongoing demand for reliable life prediction methods and understanding of materials for judging the mechanical integrity of engineering components such as turbomachinery designs drives the development of numerous sophisticated probabilistic models. 

For example in \cite{Fedelich1996,Fedelich1998}, a kinetic theory for surface microcracking processes and a balance equation describing the crack density evolution in a surface element is developed based on a Poisson Point Process description of the damage state and a Weibull distribution of initial crack lengths. Crack initiation, coalescence and propagation are covered by this theory which requires high level information on material cracking behavior however. B.H. Lee and S.B. Lee developed a stochastic model for LCF damage caused by multiaxial loading that not only considers one critical plane as proposed in \cite{Kandil1982} but also other highly loaded planes and thereby contains some sort of statistical size effect \cite{Lee2000}. A very recent probabilistic fatigue prediction model for parts under multiaxial loading is presented by Zhu \textit{et al.} who also use a critical plane approach (by \cite{FatemiSocie1988}) to derive a fatigue life distribution for a body under multiaxial load \cite{ZhuBeretta2017}.

While the previously named publications are well justified and validated for test specimens and experiment results under lab conditions, general component design requires a fast model which deals with arbitrary geometries and stress fields. Targeting these requirements, the algorithm for probabilistic LCF crack initiation prediction presented in \cite{Schmitz_Seibel2013,ASME2013Paper} is based on a local approach. It is set up as FEA postprocessor and uses the stress-field information to calculate the \textit{hazard density} for every node of the part to calculate the scale parameter of the Weibull crack initiation distribution and thus inherently considers the statistical size effect. Based on such an LCF crack initiation analysis, probabilistic fracture mechanic assessments might follow using an approach as in \cite{Amann2016,amann2016method}.

However, the life enhancing stress gradient effect was not incorporated in the algorithm of \cite{Schmitz_Seibel2013,ASME2013Paper} so far. Therefore, LCF crack initiation life predictions with very inhomogeneously loaded geometries were often too conservative \cite{Schmitz_Diss2014}. Modeling the stress gradient effect is often approached by shifting the calculated W\"ohler curve by a \textit{notch-support factors} to higher lives \cite{Harders_Roesler2007,Vormwald2007}. This principle is taken up in the model presented here but simply applying the deterministic notch support effect on top of the probabilistic prediction would still be inaccurate since it is then decoupled from the statistical size effect. The new notch support extension identifies the \textit{normalized stress gradient} at the surface, $\chi$ and uses a root function approach with two additional parameters to calculate a stress gradient dependent \textit{notch support factor} $n_\chi$. The factor $n_\chi(\chi(\mathbf{x}))$ is calculated in locally at the surface and implicitly used to reduce the local hazard density values. With this procedure, the notch support extension is well aligned with the local probabilistic approach and all shifts of the probabilistic life curves are combined shifts of by the notch effect and size effect. Estimating the notch support factor as a function of the normalized stress gradient was first proposed by \cite{Siebel1955}. Considering the experimental observations for several steels and non \ce{Fe}-alloys, Siebel \textit{et al.} have also proposed that a root function is used to model $n_\chi(\chi)$ which is a widely accepted consensus now \cite{Harders_Roesler2007,Vormwald2007} and is therefore similarly adapted for the notch support extension of the local probabilistic model from \cite{Schmitz_Seibel2013,ASME2013Paper}. This has the particular advantage that the geometry variability of the model is kept.

Another probabilistic approach to notch support is presented in \cite{Okeyoyin2013} and \cite{Owolabi2015}, for example where fatigue notch factors for high cycle fatigue (HCF) based on the random distribution of failure inducing defects in the material are computed. Hertel \textit{et al.} already take into account a \textit{macro support factor} which combines the statistical size effect and geometrical support effects \cite{HertelVormwald2010}. Karolczuk and Palin-Luc claim to model the stress gradient effect using a Weibull based distribution function which is, similar as in the model in \cite{Schmitz_Seibel2013,ASME2013Paper}, a result of FEA domain property summation \cite{Karolczuk2013}. However there, the stress gradient along the surface is evaluated in contrast to evaluating the general stress gradient which is addressed in the model presented here.

%First, the mechanism of LCF and its most widely used prediction model, the \textit{Coffin-Manson-Basquin} equation, is repeated in \autoref{subsec:The_LCF_allg}, followed by a brief description of the local probabilistic model for LCF in \autoref{subsec:The_LCF_locPM}. The new model extension for notch support consideration is described in \autoref{subsec:The_LCF_NSE} and the procedure for the extended model calibration (material parameter estimation) is described in \autoref{subsec:Main_Calib}.

First, a brief description of the local probabilistic model for LCF is given in \autoref{subsec:The_LCF_locPM}, followed by a description of the new model extension for local stress gradient based notch support consideration in \autoref{subsec:The_LCF_NSE}. 

The extended model is calibrated with LCF test data of standard and notched specimens for the materials \textit{IN-939} (CC) (\autoref{subsec:Main_IN939}), \textit{Rene80} (HIP) (\autoref{subsec:Main_Rene80}) and \textit{26NiCrMoV14-5} (\autoref{subsec:Main_Steel14-5}). Calibration for \textit{IN-939} additionally requires the use of temperature models for the Coffin-Manson-Basquin (CMB) parameters which are calibrated in a two step process. Subsequently the calibrated model is validated by predicting the LCF crack initiation life of cooling hole specimens, giving more accurate results than a deterministic approach. Furthermore, the model is calibrated for \textit{Rene80}(HIP) with two different data sets described in \ref{subsec:Main_Rene80}. There it is shown that the model can be robustly calibrated with data of only one notched specimen geometry. Finally in \ref{subsec:Main_Steel14-5} it is shown that the model can also be successfully applied to the class of martensitic steels. In all cases presented, the prediction confidence is assessed by fitting bootstrap curves and calculating the percentiles.

%% Liu & Mahadevan
%Amongst others, Liu and Mahadevan discuss the concept of probabilistic \textit{S-N}-curves in combination with fatigue damage accumulation and validate test data results of six different material classes \cite{Liu2006}.

%% file: Theory.tex
\section{Probabilistic LCF and notch support}\label{sec:The_LCF}
%%The mechanism of LCF in polycrystalline metals is briefly recapped in \autoref{subsec:The_LCF_allg} followed by a repetition of the key ingredients of the local probabilistic model in \autoref{subsec:The_LCF_locPM} which is used as basis for the notch support extension presented in \autoref{subsec:The_LCF_NSE}.
%In the following \autoref{subsec:The_LCF_locPM}, the key ingredients of the local probabilistic model are briefly recapped as they are the basis for the notch support extension presented in \autoref{subsec:The_LCF_NSE}.
%
%%\subsection{The Low Cycle Fatigue mechanism in polycrystalline metals}\label{subsec:The_LCF_allg}
%%As the later discussed notch support effect directly influences the mesoscopic processes leading to surface crack initiation, they are repeated here once more. Macroscopic body deformation increases the crystal dislocations density within the metal grains. While dislocations are first uniformly distributed across the grain volume, cyclic loading then causes them to move within the crystal and eventually accumulate and persist forming \textit{persistent slip bands} (PSB) thereby activating shear deformation at the crystal slip planes. The maximum intergranular deformation which might eventually lead to a crack occurs in grains whose PSB's are oriented \unit{45}{\degree} to the direction of external load in accordance with \textit{Schmid's shear stress law} \cite{Schmid_Boas1950}. Finally, PSB's in surface grains dislocate, forming intrusions and extrusions which then concentrate the stress and hence initiate fatigue micro cracks within the grain. 

\subsection{Local Probabilistic Model for LCF}\label{subsec:The_LCF_locPM}
The local probabilistic model for LCF discussed here was introduced in \cite{Schmitz_Seibel2013} and \cite{ASME2013Paper}. One of its major advantages is the ability to account for the statistical size effect which means that a part with larger loaded surface area is more likely to show LCF crack initiation than a part with smaller loaded surface area. The other large benefit of the local probabilistic model is the delivery of geometry independent CMB parameters after calibration to test points.

In deterministic methods, the \textit{Coffin-Manson-Basquin} (CMB) equation is widely used to model the relationship between maximum (elasto-plastic) strain $\varepsilon_{a}$ in the component and its crack initiation life $N_i$ \cite{Coffin1954,Coffin1974,Basquin1910} as it is well justified by the physical processes at mesoscopic scale and their stochastic nature \cite{Sornette1992}.
\begin{equation}
\varepsilon_a=\frac{\sigma_f^\prime}{E}\cdot(2N_i)^b+\varepsilon_f^\prime\cdot(2N_i)^c\label{eq:The_CMBeq}
\end{equation}
\ref{eq:The_CMBeq} is parametrized by the cyclic Youngs's modulus $E$, the \textit{fatigue strength} coefficient and exponent $\sigma_f,\ b$ and the \textit{fatigue ductility} coefficient and exponent $\varepsilon_f,\ c$ which are generally material and geometry dependent. Hence for deterministic LCF life prediction (usually crack initiation life $N_i$) the CMB parameters are first determined by fitting the CMB model to LCF test points and \ref{eq:The_CMBeq} is numerically inverted for a given maximum component strain amplitude $\varepsilon_a$. For component design however, the resulting value $N_i$ is not a final result but usually evaluated considering safety factors to accommodate for size effects. A direct approach to the statistical size effect is incorporated in the following local probabilistic model.

In contrast to the deterministic approach, the number of cycles to crack initiation $N_i$ is considered as a random variable which is Weibull distributed with the cumulative distribution function
\begin{align}
&F_N(n)=1-\mathrm{exp}\left(-\left(\frac{n}{\eta}\right)^m\right)\label{eq:The_WeibFn}
\end{align}
where $m$ and $\eta$ are the Weibull shape and scale parameter. LCF cracks initiate at the surface and only influence stress fields on micro- and mesoscales which infers that crack formation processes in different locations are stochastically independent \cite{Gottschalk_Schmitz2014}. Hence, hazard rates $h_j(n)$ for crack initiation at the surface patches $\{A_j\}_{j=1...k}$ are calculated as integrals over a function depending on the local elasto-plastic strain
\begin{align}
h(n)=\int_{\partial\Omega}{\varrho(n;\varepsilon_a(\mathbf{x}),T(\mathbf{x}))\,dA}.\label{eq:The_sum_h}
\end{align}
The integrand $\varrho(n;\varepsilon_a(\mathbf{x}),T(\mathbf{x}))$ is the \textit{hazard density}. With the \textit{cumulative hazard rate} $H(n)=\int_{\partial\Omega}{h(s)\,ds}$ and its relationship to \ref{eq:The_WeibFn}, $F_N(n)=1-\mathrm{exp}\left(-H(n)\right)$, the hazard density can be derived to 
\begin{align}
\varrho(n;\varepsilon_a,T)=\frac{m}{N_{i_\mathrm{det}}(\varepsilon_a,T)}\left(\frac{n}{N_{i_\mathrm{det}}(\varepsilon_a,T)}\right)^{m-1}.\label{eq:The_rho}
\end{align}
%$N_i(\mathbf(x))$, the deterministic crack initiation life for the unit specimen, is assumed to be equal to the Weibull scale parameter. It is calculated by numerically inverting the CMB equation 
%\begin{align}
%\varepsilon_a(\mathbf{x},T(\mathbf{x}))=\frac{\sigma^\prime_f}{E}\cdot(2N_i\left(\mathbf{x})\right)^b+\varepsilon^\prime_f\cdot\left(2N_i(\mathbf{x})\right)^c\label{eq:The_CMBeq}
%\end{align}
%whose parameter set is then size- and geometry-independent \cite{Schmitz_Seibel2013}. With \ref{eq:The_rho}, the final expression for the Weibull scale parameter $\eta$ of the Weibull distribution function \eqref{eq:The_WeibFn} is found to be
The final expression for the Weibull scale parameter $\eta$ of the Weibull distribution function \eqref{eq:The_WeibFn} is found to be
\begin{align}
\eta=\left(\int_{\partial\Omega}{\frac{1}{N_{i_\mathrm{det}}\left(\varepsilon_a(\mathbf{x}),T(\mathbf{x})\right)^m}dA}\right)^{-1/m}.\label{eq:The_ScaleSurfInt}
\end{align}
By integrating over all hazard density values of a body's surface, the scale parameter $\eta$ inherently considers the statistical size effect. Thereby the model is able to predict the median life $N_i$ of an arbitrarily shaped body within the limits of stress modeling by continuum mechanics. In order to compare the effect of different critical sizes a \textit{size effect factor} $SE$ is calculated for every component evaluation
\begin{align}
SE&=\frac{N_{i_\mathrm{comp}}}{N_{i_\mathrm{smooth}}}.\label{eq:Main_SizeEff}
\end{align}
In \ref{eq:Main_SizeEff} $N_{i_\mathrm{comp/smooth}}$ are the probabilistic life numbers of the assessed component and a comparably loaded, standardized smooth specimen. Often engineering parts with inhomogeneous geometry have stresses concentrated to small areas and hence low deterministic life in at the respective integration point whereas the other areas are far less critical. Given the risks in the component are sufficiently confined locally, the integration in \ref{eq:The_ScaleSurfInt} results in higher probabilistic life compared to evaluating a standard LCF specimen for the load situation of the lowest life point in the component. Hence $SE>1$ which means the W\"ohler curve corresponding to the engineering part is shifted along the $N_i$-axis to higher life.

%Using of the CMB equation \eqref{eq:The_CMBeq} in \ref{eq:The_ScaleSurfInt} links the model to the actual mesoscale processes of LCF crack initiation \cite{Sornette1992}.
For an analysis with the local probabilistic model, the local strain values $\varepsilon_a(\mathbf{x},T(\mathbf{x}))$ of the assessed body, e.g. a test specimen, have to be determined as input for \ref{eq:The_CMBeq}. This is done by retrieving the nodal von Mises stress result of a linear elastic FEA model and applying it to a shakedown function $SD^{-1}:\sigma^{el}_{v}\rightarrow\sigma_{v}$, e.g. the Neuber rule. Strain values are then derived by subsequently applying the Ramberg-Osgood equation to the elasto-plastic stresses.
%\begin{equation}
%\frac{(K_t\sigma^{el}_{v})^2}{E}=\frac{\sigma_{v}}{E}+\sigma_{v}\left(\frac{\sigma_{v}}{K}\right)^{1/n}\label{eq:The_NeuberSDeq}
%\end{equation}
%and subsequently the applying the Ramberg-Osgood equation to the elasto-plastic stresses $\sigma_{v}$
%\begin{equation}
%\varepsilon_a=\frac{\sigma_{v}}{E}+\left(\frac{\sigma_{v}}{K}\right)^{1/n}.\label{eq:The_ROeq}
%\end{equation}
With this setup as an FEA postprocessor \cite{Schmitz_LCF7Paper}, the local probabilistic model can be applied in a very straight forward way to assess the LCF life of bodies with arbitrary geometry.

\subsection{Notch Support Effect}\label{subsec:The_LCF_NSE}
This subsection recapitulates the basics of the notch support effect and how the probabilistic model for LCF is extended to account for it.

It is well known that irregular shaped parts with local stress maxima show higher LCF life than predicted by considering homogenous nominal stresses in the carve or notch root only. The reason for that is the \textit{stress gradient-} or \textit{notch support effect} which means that stress levels dropping from the surface into the bulk of the part inhibit LCF crack initiation in the critical grain. This is well comprehensible when keeping in mind that the movement and time to accumulation of crystal dislocations is governed by the lattice strain \cite{Sornette1992}. Furthermore, stresses dropping across the diameter of a critical surface grain retard the formation of persistent slip bands and hence macroscopic crack initiation. In other words, the lower stress regions underneath the surface support the material with inhomogeneous stress distribution.

\begin{figure}[t]
\begin{minipage}[t]{0.49\columnwidth}
	\centering
	\includegraphics[width=\textwidth]{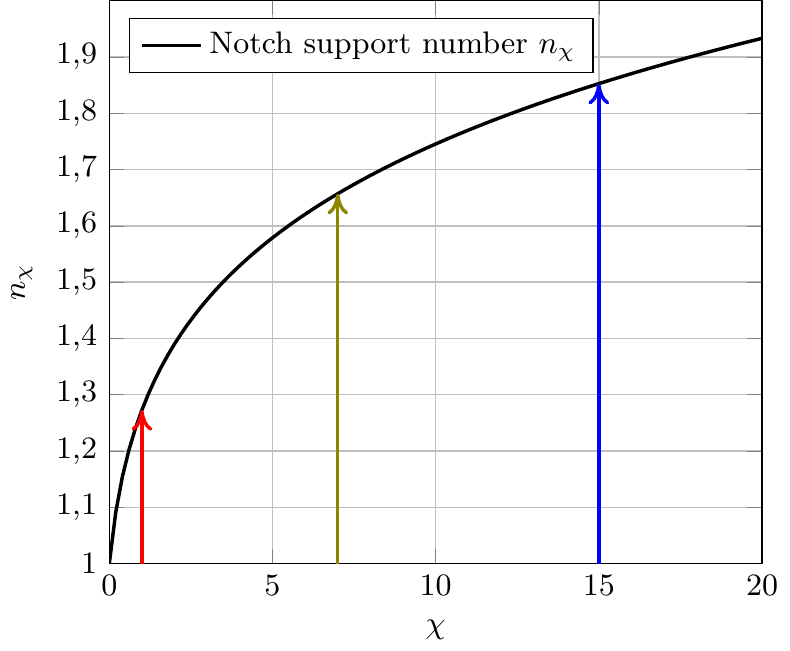}
	\end{minipage}
\begin{minipage}[t]{0.49\columnwidth}
	\centering
	\includegraphics[width=\textwidth]{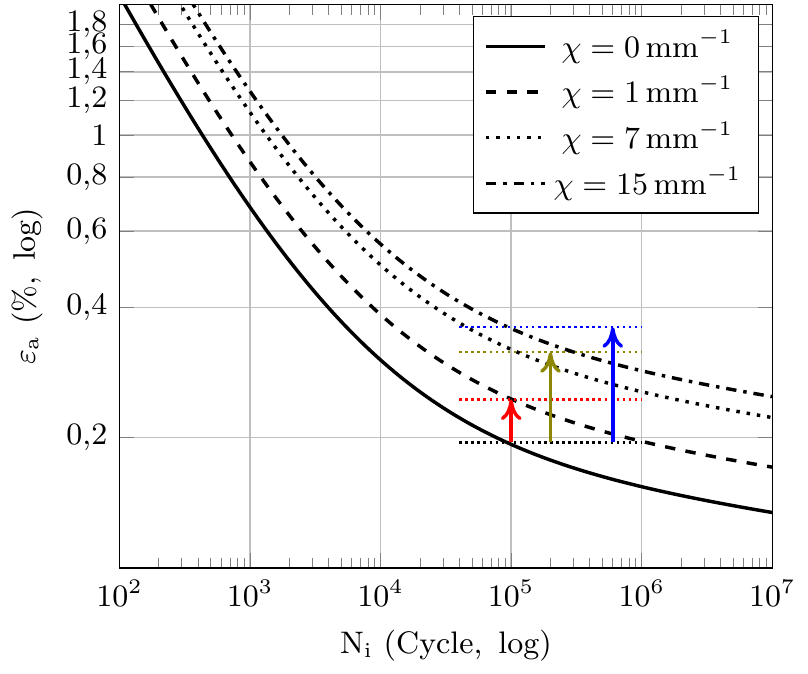}
\end{minipage}

\caption[]{The left image shows a typical dependency of $n_\chi$ on $\chi$ as observed for most ferrous and non ferrous steels \cite{Harders_Roesler2007,Siebel1955}. Exemplary CMB W\"ohler curves and their shift due to notch support are shown in the right image. Arrows correspond to each other in both images.}
\label{fig:The_ChiVisualisierung}
\end{figure}

In order to use LCF life prediction models such as the CMB equation, the load magnitude there is decreased by a notch support factor $n_\chi$ which is a function of the normalized stress gradient $\chi$ at the surface $\partial\Omega$.
\begin{align}
\chi(\mathbf{x})&=\frac{1}{\sigma^{el}_{v}\mathbf(\mathbf{x})}\cdot\nabla\sigma^{el}_{v}(\mathbf{x})\, &\text{with}\ \mathbf{x}\in\partial\Omega.\label{eq:The_chiEq}
\end{align}
While \cite{Siebel1955} used a one parametric root function
\begin{equation}
n_\chi=1+\sqrt{s_g\cdot\chi}
\end{equation}
where $s_g$ is the distance of persistent slip bands in the failing grain, here, a similar but two-parametric approach with parameters $A$ and $k$ is here chosen for calculating $n_\chi$. Deterministically, the maximum strain value $\varepsilon_a$ in the CMB equation is replaced with $\varepsilon_a/n_\chi$. Since $n_\chi\geq 1$ for all $\chi\geq 0$, using $\varepsilon_a/n_\chi$ shifts W\"ohler curves along the $\varepsilon_a$-axis to higher strain values as shown in \autoref{fig:The_ChiVisualisierung}. This is approximately modeling the dependency of the shift in cyclic fatigue strength (or strain) on $\chi$ which is observed for differently shaped notched specimens at a certain number of load cycles.

%\begin{figure}[htbp]
	%\centering
	%\input{Images/Theory/Chi_Visualisation1.tex}
	%\caption[]{Typical dependency of $n_\chi$ on $\chi$ as observed for most ferrous and non ferrous steels \cite{Harders_Roesler2007,Siebel1955}.}
	%\label{fig:The_ChiVisualisierung1}
%\end{figure}
%\begin{figure}[htbp]
	%\centering
	%\input{Images/Theory/Chi_Visualisation2.tex}
	%\caption[]{Exemplary CMB W\"ohler curves and their shift due to notch support. Arrows correspond to those in \ref{fig:The_ChiVisualisierung1}.}
	%\label{fig:The_ChiVisualisierung2}
%\end{figure}

In the extended probabilistic model for LCF, all local strain values $\varepsilon_a(\mathbf{x})$ are divided by the respective local notch support factor $n_\chi(\mathbf{x})$ at every integration point. So the values $N_{i_\mathrm{det}}(\varepsilon_a(\mathbf{x}),T(\mathbf{x}))$ in \ref{eq:The_rho} are calculated by inverting
\begin{equation}
\frac{\varepsilon_a(\mathbf{x})}{n_\chi\left(\chi(\mathbf{x})\right)}=\frac{\sigma^\prime_f}{E}\cdot(2N_{i_\mathrm{det}})^b+\varepsilon^\prime_f\cdot(2N_{i_\mathrm{det}})^c \label{eq:The_CMBwithNoS}.
\end{equation}
Using \ref{eq:The_CMBwithNoS} leads to probabilistic W\"ohler curves shifted along the the $\varepsilon_a$ as well but additionally, they are also shifted to horizontally to higher $N_i$ by the statistical size effect. The size effect is usually quite significant for parts with confined ares of concentrated stresses and should therefore always be considered.

The use of the $\chi$-based notch support factor keeps the flexibility of the local probabilistic model to assess arbitrary geometries since $\chi(\mathbf{x})$ is also identified from the nodal linear-elastic stress values of FEA results. 

The derivative $\nabla\sigma^{el}_{v}(\mathbf{x})$ is computed by applying the chain rule to the von Mises stress equation and finally calculating the component stress derivatives using FEA shape functions which interpolate stress from the surface to the first interior layer of finite elements. The latter are hard coded for most element types, such that computing \ref{eq:The_chiEq} is fast and precise. This is also necessary, because instead of evaluating the notch support effect only at the location of highest load it is evaluated at every integration point at the surface once the the nodal $\chi(\mathbf{x})$ values are identified. This follows the philosophy of local contributions to the overall hazard rate, hence enables the model to use geometry independent parameters and directly accounts for the combined size- and notch support effect. However, this requires, that FEA models of all notched specimen test points have to be solved and their nodal stress tensor and temperature results have to be extracted once in order to calculate the integral \eqref{eq:The_ScaleSurfInt}, numerically. The same procedure is necessary for calculating the curve points of the predicted probabilistic W\"ohler curves and is hence computationally more costly than the probabilistic LCF model calibration and curve prediction in \cite{Schmitz_Seibel2013} with smooth specimen data only permitting an analytical solution of \ref{eq:The_ScaleSurfInt}.

In order to estimate $A$ and $k$ for the notch support extension in the deterministic model, test data of at least \textit{two differently notched} specimen sets has to be acquired. The strain amplitude ratio in the LCF W\"ohler curves of notched and smooth specimen data at a certain cycle number is plotted versus the normalized stress gradient $\chi$ in the notch root. Having done that for several notch specimen types, the model for $n_\chi(\chi)$ is fitted to those points which creates a curve such as the left plot in \autoref{fig:The_ChiVisualisierung}. Improving the data basis for such a fit requires more test points at different $\chi$ values which creates the effort to manufacture and test specimens of additional geometries. In contrast, the calibration of the extended local probabilistic model for LCF requires only tests of one notch geometry are required \footnote{More data sets for calibrating and validating the results are preferable of course}.

%% file: Experiments.tex
\section{Calibration and Validation of the probabilistic LCF/notch support model}\label{sec:Main}
The probabilistic LCF model considering the notch support effect requires seven material parameters $\sigma^\prime_f,\ b,\ \varepsilon^\prime_f,\ c,\ A$ and $k$ which are generally temperature dependent. But in this work, temperature models are only introduced for the CMB parameters while notch support parameters $A,\ k$ and the Weibull shape parameter $m$ assumed to be constant in temperature. 

In order to calibrate the model parameters, all LCF data points have to be fitted simultaneously because in the local probabilistic model the Weibull shape parameter accounts for distribution scatter and the statistical size effect which causes a shift of the probabilistic median life. Ultimately, the shift in crack initiation life of non-smooth specimens such as notched specimens is always a combination of notch support effect and the size effect. Therefore, the consecutive estimation of only the CMB parameters first and the separate estimation of the notch support parameters in a second step, like the deterministic method, is inaccurate. Particularly $A,\ k$ and $m$ need to be estimated simultaneously. 

Since it is able to handle censored data, the \textit{maximum likelihood estimation} (MLE) method is used to calibrate the local probabilistic model. There, the negative log-likelihood sum is minimized using the Nelder-Mead algorithm, due to the complex numerical operations involved, 

In the following subsections the extended local probabilistic model is calibrated for the two superalloys \textit{Rene80} and \textit{IN-939} and the steel \textit{26NiCrMoV 14-5} for validating the predictive power of the notch support model and its applicability to different material classes. These materials are often used for turbomachinery components.

\subsection{\textit{IN-939} (CC)}\label{subsec:Main_IN939}
The first case considers \textit{IN-939}, a polycrystalline \ce{Ni}-based superalloy. Smooth and notch specimen data is used for model calibration and the median life of cooling hole specimens is predicted which also required the calibration of a temperature model for the CMB parameters. The probabilistic LCF life evaluation of cooling hole geometries is very interesting since \textit{IN-939} is typically used for casting gas turbine blades and vanes. Its average chemical composition is given in \autoref{tab:ChemCompIN939}

\begin{table}[htbp!]
\centering% NICHT \begin{center}
\begin{tabular}{|c|c|c|c|c|c|c|c|c|c|c|}%\label{table.1.1}
  \hline
  Element & Ni & Cr & Co & Ti & Mo & W & Al & C & B & Zr \\
  \hline
  Weight - $\%$ & bal. & 22.5 & 19.0 & 3.7 & 2.0 & 2.0 & 1.9 & 0.15 & 0.005 & 0.025\\
  \hline
\end{tabular}
\caption{Chemical composition of \textit{IN-939}. The examined specimens were cast and heat treated such that the material had a yield strength $YS_{0.2}\geq\unit{704}{\mega\pascal}$ at \unit{400}{\celsius}.}\label{tab:ChemCompIN939}
\end{table}

\begin{figure}[htbp]
\begin{minipage}[htbp]{0.49\columnwidth}
	\centering
	\includegraphics[width=\textwidth]{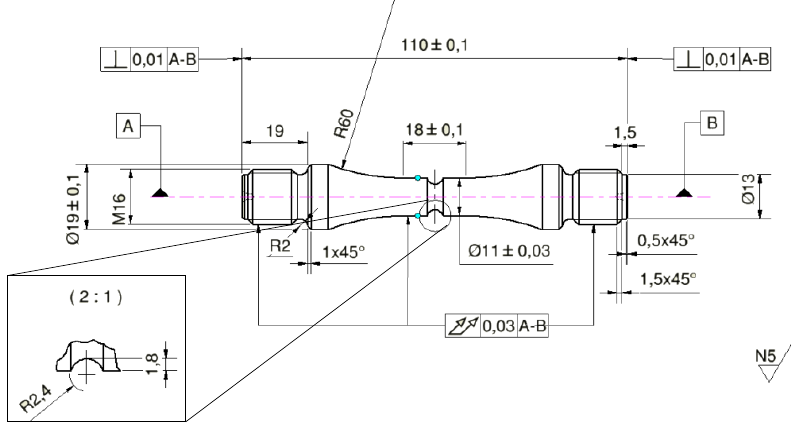}
\end{minipage}
\begin{minipage}[htpb]{0.49\columnwidth}
	\centering
	\includegraphics[width=\textwidth]{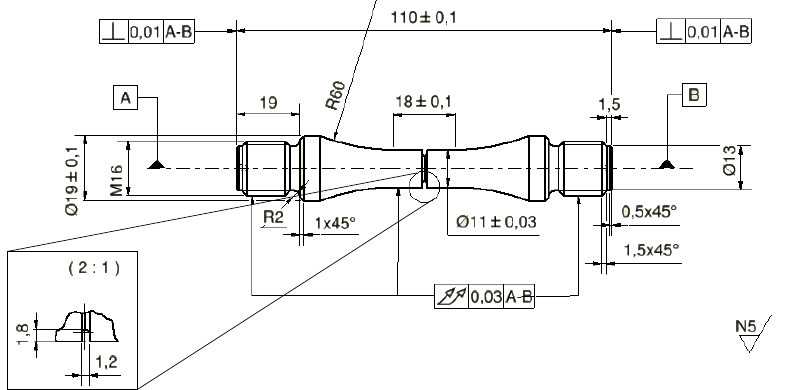}
\end{minipage}
\caption[]{Geometries of the notch specimen type used for test data generation as used in the \textit{Rene80} (HIP) test campaign \cite{Beck_Gottschalk2014}. The left sketch shows a notched specimen with low stress concentration ($K_t=1.62$), while the sharply notched specimen in the right sketch causes higher stress concentration ($K_t=2.6$). Similar specimen geometries were used for \textit{IN-939} (CC) tests.}
	\label{fig:Main_IN939_NotchSketch}
\end{figure}

LCF data of five different specimen geometries is examined. The standard specimens have a homogenous, cylindrical gauge section with $r_\mathrm{Gauge}=\unit{3.175}{\milli\meter}$. The two notch specimen types have circumferential notch with radii of \unit{9.13}{\milli\meter} and \unit{3.13}{\milli\meter} respectively in the middle of the gauge section and a notch root radius of also \unit{3.175}{\milli\meter}. \autoref{fig:Main_IN939_NotchSketch} shows geometry sketches of this specimen type as it was used in the \textit{Rene80} test campaigns. This specimen type was used for notch specimen tests of all materials but with slightly similar dimensions.

%\begin{figure}[htbp!]
	%\centering
%\input{Images/Main/Rene80/large_Notch_sketch.tex}
%\input{Images/Main/Rene80/small_Notch_sketch.tex}		
%\caption[]{Geometries of the notch specimen type used for test data generation as used in the \textit{Rene80}""(HIP) test campaign \cite{Beck_Gottschalk2014}. The upper sketch shows a notched sample with low stress concentration ($K_t=1.62$), while the sharply notched specimen in the sketch below causes higher stress concentration ($K_t=2.6$). Similar samples were used for \textit{IN-939}""(CC) tests.}
	%\label{fig:Main_IN939_NotchSketch}
%\end{figure}

\begin{figure}[htbp]
\begin{minipage}[t]{0.49\columnwidth}
	\centering
\fbox{\includegraphics[width=0.98\textwidth]{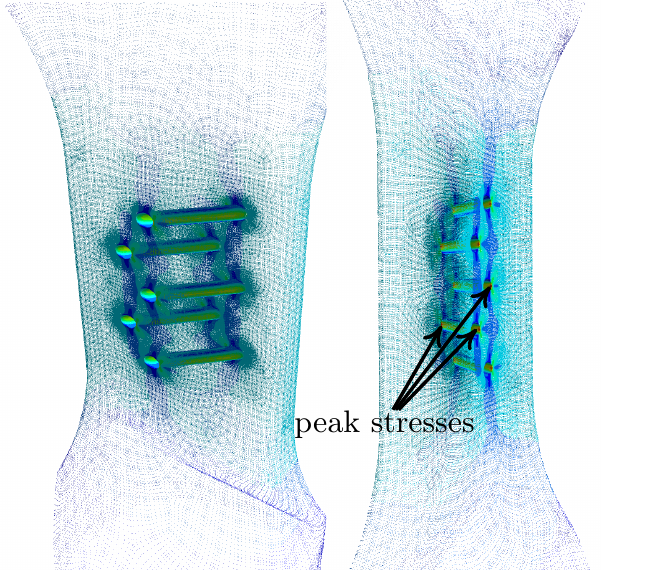}}
\end{minipage}
\begin{minipage}[t]{0.49\columnwidth}
	\centering
\fbox{\includegraphics[width=0.98\textwidth]{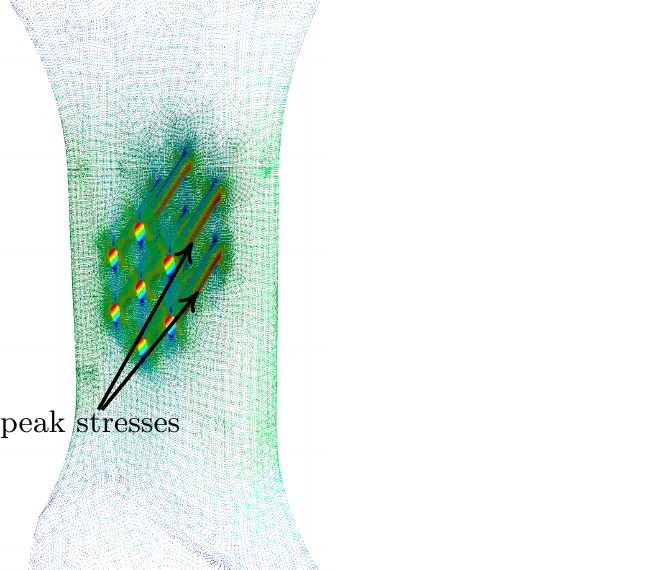}}
\end{minipage}
\caption[]{The cooling hole specimen geometry \textit{CH-5} with 5 holes leading to a stress concentration with $K_t=4.0$ is shown in  the left box. Areas of high stresses are very confined to the transition area between cooling hole channel and specimen flank at the sharp angle side. The cooling hole specimen geometry \textit{CH-7} with 7 holes leading to a stress concentration with $K_t=2.1$. Areas of high stresses extend elongated along the entire sides of cooling hole channels. As a consequence, the critical area in the specimen is higher and therefore a lower size effect shift is expected.}
\label{fig:Main_CH_surfStress}
\end{figure}

Additionally to those, two types of cooling hole specimens with the same strength class were tested in order to obtain more information on LCF crack initiation of turbine blade or vane cooling holes. \autoref{fig:Main_CH_surfStress} shows scatter plots of the elasto-plastic stress distributions at the free surfaces of the specimen type with $K_t=4.0$ (cooling hole channels horizontally) and $K_t=2.1$ (cooling hole channels inclined by \unit{45}{\degree}) respectively.

%\begin{figure}[htbp!]
	%\centering
%\input{Images/Main/IN939/M5003_CHALT_SurfStress.tex}
%\caption[]{Cooling hole specimen with 5 holes leading to a stress concentration with $K_t=4.0$. Areas of high stresses are very confined to the transition area between cooling hole channel and specimen flank at the sharp angle side.}
	%\label{fig:Main_CHALT_surfStress}
%\end{figure}
%\begin{figure}[htbp!]
	%\centering
%\input{Images/Main/IN939/M5003_CH45_SurfStress.tex}
%\caption[]{Cooling hole specimen with 7 holes leading to a stress concentration with $K_t=2.1$. Areas of high stresses extend elongated along the entire sides of cooling hole channels. As a consequence, the critical area in the specimen is higher and therefore a lower size effect shift is expected.}
	%\label{fig:Main_CH45_surfStress}
%\end{figure}

All test specimens used for calibration and validation here underwent heat treatments resulting in the same strength class. The specimen type and test environment of the available LCF test data\footnote{Siemens AG proprietary test results.} is listed in \autoref{tab:IN939_TestCollection}.
\begin{table}[htbp!]
\centering% NICHT \begin{center}
\begin{tabular}{|p{0.54\textwidth}|p{0.4\textwidth}|}
  \hline
  Geometry & Test temperatures, $T_\text{Test}$\\
  \hline \hline
  smooth cylindrical $r_\text{Gauge}=\unit{3.175}{\milli\meter}$, \textit{Smooth}& \unit{23}{\celsius}, \unit{427}{\celsius}, \unit{760}{\celsius}, \unit{800}{\celsius}, \unit{850}{\celsius}, \unit{871}{\celsius}, \unit{900}{\celsius}, \unit{1000}{\celsius}\\
	\hline
	Notched with $r_\text{Notch}=\unit{0.913}{\milli\meter}$, $K_t=2$, \textit{NS-1} & \unit{500}{\celsius}, \unit{750}{\celsius}, \unit{850}{\celsius}, \unit{950}{\celsius}\\
  \hline
	Notched with $r_\text{Notch}=\unit{0.313}{\milli\meter}$, $K_t=3$, \textit{NS-2} & \unit{500}{\celsius}, \unit{750}{\celsius}, \unit{850}{\celsius}, \unit{950}{\celsius}\\
  \hline
	Cooling hole, 7 holes, $r_\mathrm{CH}=\unit{0.35}{\milli\meter}$, $K_t=2.1$, \textit{CH-7} & \unit{649}{\celsius}, \unit{871}{\celsius}\\
	\hline
	Cooling hole, 5 holes, $r_\mathrm{CH}=\unit{0.35}{\milli\meter}$, $K_t=4.0$, \textit{CH-5} & \unit{649}{\celsius}, \unit{871}{\celsius}\\
  \hline
\end{tabular}
\caption{Available LCF tests for \textit{IN-939}.}
\label{tab:IN939_TestCollection}
\end{table}

\begin{figure}[htbp]
\begin{minipage}[t]{0.50\columnwidth}
	\centering
\includegraphics[width=\textwidth]{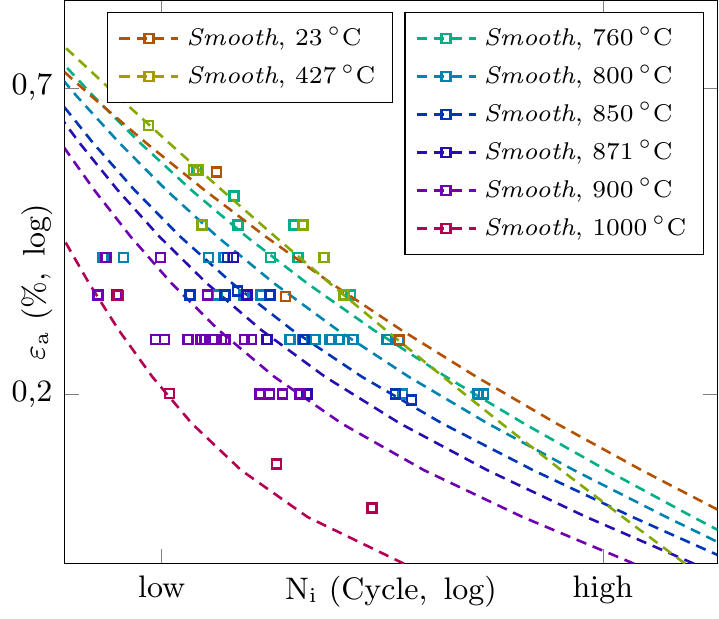}
\end{minipage}
\begin{minipage}[t]{0.50\columnwidth}
	\centering
\includegraphics[width=\textwidth]{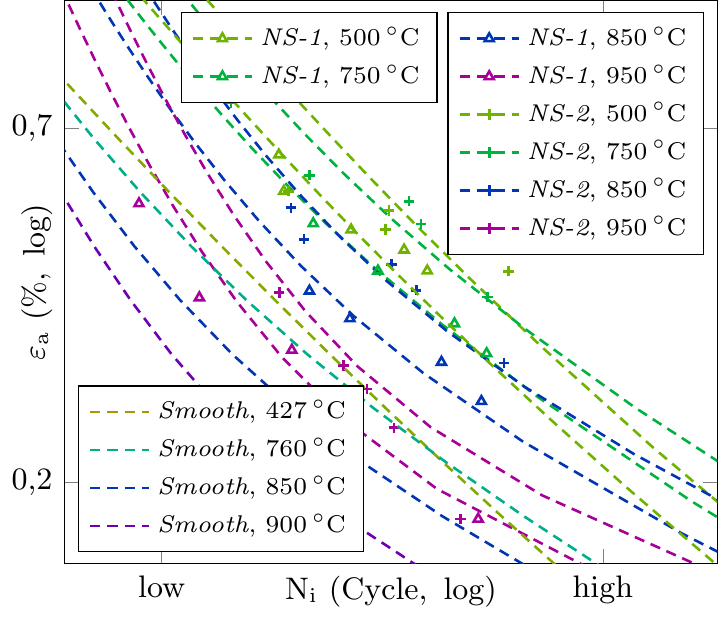}
\end{minipage}
\caption[]{Prob. W\"ohler curves and LCF test data points of \textit{IN-939} (CC) smooth specimens are shown in the left plot. All curves are fit curves. The CMB-parameters are modeled temperature dependent using polynomial temperature model functions. The respective zero'th order parameter is optimized simultaneously with notch support parameters and Weibull shape. The right plot shows the notch specimen data points and their corresponding fit curves.}
\label{fig:Main_M5003_clean1018}
\end{figure}

Temperature model functions for the CMB parameters are determined by estimating the CMB coefficients for each individual smooth specimen data set first\footnote{This was done applying a Siemens proprietary curve fitting tool for smooth specimen data which fits probabilistic \textit{E-N}-curves with Maximum Likelihood estimation as well, but neglects the statistical size effect.} (shown in the let plot in \autoref{fig:Main_M5003_clean1018}) and secondly fitting polynomial ansatz functions to those temperature-individual CMB parameters. The 0th order parameters of these polynomials
\begin{align}
b(t)&=\mathbf{b_0}+b_1\cdot T +...\notag\\
c(t)&=\mathbf{c_0}+c_1\cdot T +...\notag\\
\sigma^\prime_f(T)&=\boldsymbol{\sigma}_\mathbf{f0}+\sigma_{f1}\cdot ...\notag\\
\varepsilon^\prime_f(T)&=\boldsymbol{\varepsilon}_\mathbf{f0}+\epsilon_{f1}\cdot ...\notag
\end{align}
which determine the absolute level of the parameter curve are later varied in the main MLE procedure and estimated simultaneously with $A\, k$ and $m$. The second step then uses the notch specimen data points shown in the right plot of \autoref{fig:Main_M5003_clean1018}. Using such semi-hard coded dependencies, the temperature trend was held fixed while the absolute of the CMB parameters was adapted to accommodate the combination of fatigue-strength and ductility influence and the size effect on the \textit{E-N}-curve position. However, note that temperature model functions are only used here to calculate the precise CMB parameters at the temperatures at which test points are available. 
%A component LCF life assessment does not use temperature model functions but tabulated values are interpolated linearly and logarithmically in order to cover all occurring temperatures.
%Since the LCF crack initiation probability of the notched specimens is subject to a significant size effect, the Weibull shape decreased in the second MLE fit compared to the pre-fit result. 

%\begin{figure*}[htbp!]
	%\centering
%%\input{Images/Main/IN939/M5003_cleanWoehlerPlot_StandKt2Kt3_20170411_1018.tex}
%\includegraphics[width=\textwidth]{figures_for_Submission/M5003_cleanWoehlerPlot_StandKt2Kt3_20170411_1018.pdf}
%\caption[]{Prob. W\"ohler curves and LCF test data points of \textit{IN-939}""(CC). Test data points of both notch specimen geometries (leading to $K_t=2$ and $K_t=3$) are used for fitting. Only CMB-parameters are assumed temperature dependent using polynomial temperature model functions. The respective zero'th order parameter is optimized simultaneously with notch support parameters and Weibull shape.}
	%\label{fig:Main_M5003_clean1018}
%\end{figure*}

The dashed lines are the probabilistic median-W\"ohler curves for LCF crack initiation life distribution depending on maximum surface strain. Test points are not necessarily crossed by this curve but are found in the scatter band of the density distribution function for $N_i$. The left plot in \autoref{fig:Main_M5003_clean1018} also shows very distinctively, how test points as well as their corresponding W\"ohler curves are shifted to lower crack initiation life with increasing temperature due to decreased fatigue strength. Additionally, the right plot in \autoref{fig:Main_M5003_clean1018} shows that notched specimens \textit{NS-2} ($K_t=3$) have higher crack initiation life than the \textit{NS-1} specimens ($K_t=2$) due to the higher stress gradient under the surface in the notch root emerging from the sharper notch geometry.

%\begin{figure}[htbp!]
	%\centering
%\input{Images/Main/IN939/M5003_MedDistFun_IN939Kt2_20170411_1018.tex}
%\caption[]{Test points and predicted probabilistic W\"ohler curve for the $K_t=2$ data set at \unit{950}{\celsius}. The Weibull density distribution functions are shown for every tested strain level and the position of the \unit{50}{\%}-quantile is marked.}
	%\label{fig:Main_M5003_DistFun_Examp}
%\end{figure}
\begin{figure}[htbp]
%\begin{minipage}[htb]{0.5\columnwidth}
	\centering
	\includegraphics[width=0.5\textwidth]{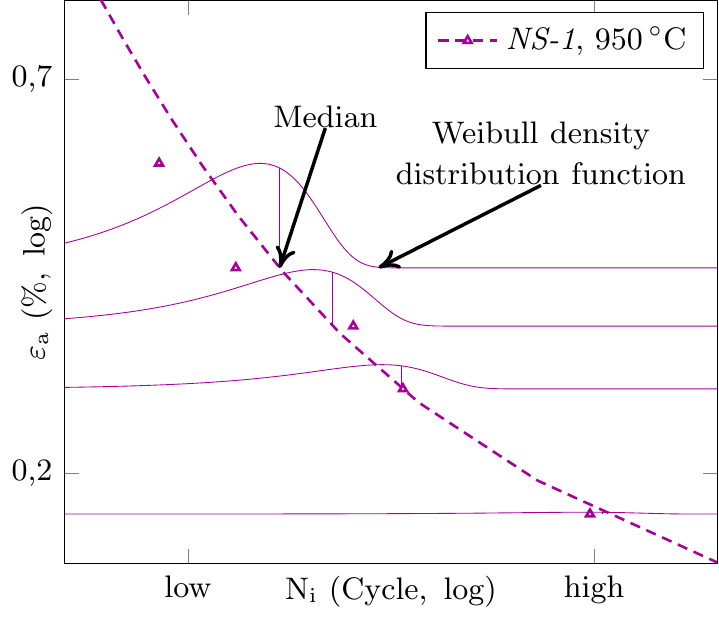}
	%\end{minipage}
	\caption[]{Test points and predicted probabilistic W\"ohler curve for the \textit{NS-1} specimen data set at \unit{950}{\celsius}. At every tested strain level the Weibull density distribution function and the position of the \unit{50}{\%}-quantile is shown.}
	\label{fig:Main_M5003_DistFun_Examp}
\end{figure}

\autoref{fig:Main_M5003_DistFun_Examp} shows an example of the predicted W\"ohler curve for the \unit{950}{\celsius} data set of the notched geometry \textit{NS-1} and the density distribution functions for every respective data point.

\begin{figure}[htbp]
	\includegraphics[width=\textwidth]{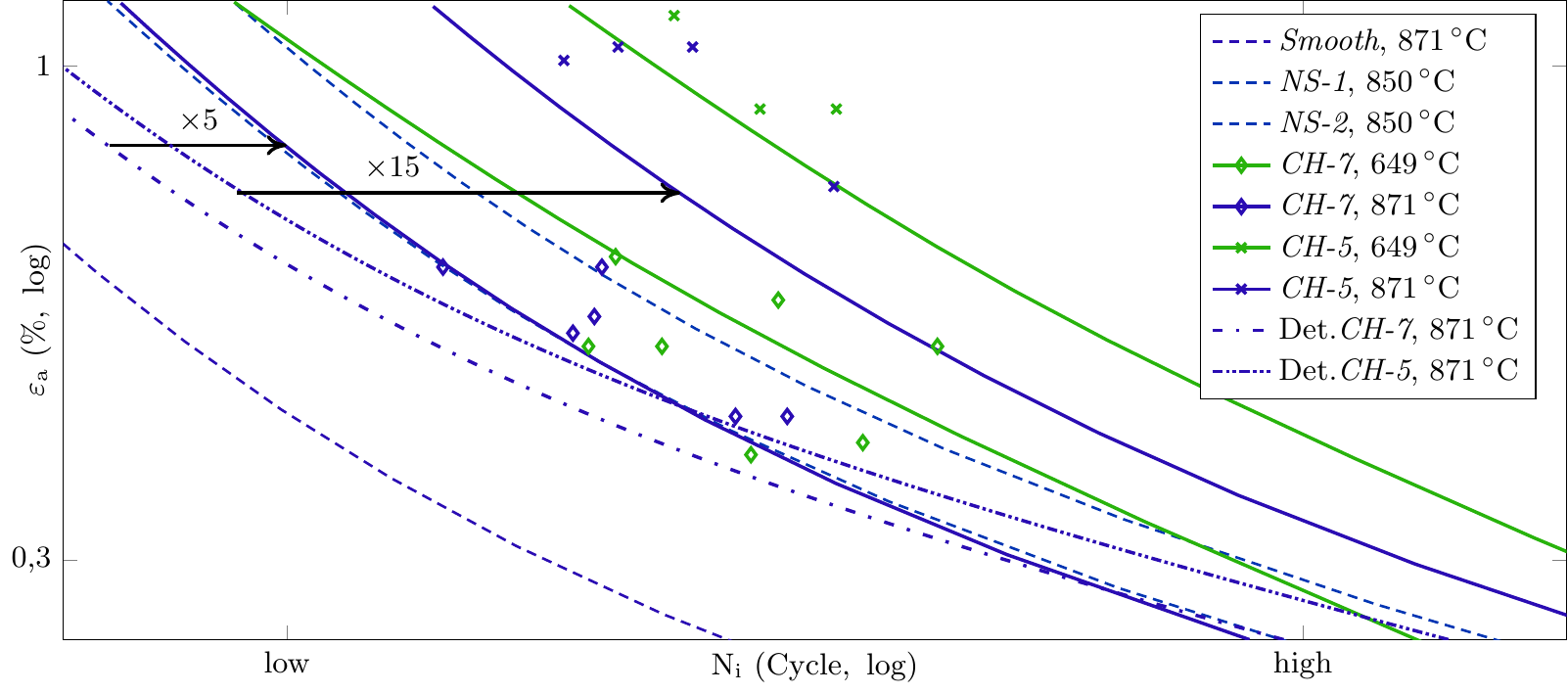}
	\caption[]{Prob. W\"ohler curves for \textit{IN-939} (CC) specimens at \unit{649}{\celsius} and \unit{871}{\celsius}. Dashed lines represent fit curves of smooth and notched specimens. Solid lines represent prob. life predictions curves of the cooling hole specimen geometries \textit{CH-5} (left in \ref{fig:Main_CH_surfStress}) and \textit{CH-7} (right in \ref{fig:Main_CH_surfStress}). Dash-dotted lines represent deterministic life predictions of cooling hole specimens at \unit{871}{\celsius}.}
	\label{fig:Main_M5003_clean1018_CH_Detcomp}
\end{figure}

Now the most important validation case of this work is shown and described. The life predictions for the two different cooling hole specimens with approximate stress concentration factors of $K_t=4.0$ and $K_t=2.1$ are presented. \autoref{fig:Main_M5003_clean1018_CH_Detcomp} shows a comparison between the deterministic W\"ohler curve predictions considering notch support using the inverse of 
\begin{align}
\frac{\varepsilon_a\left(\mathbf{x}_\mathrm{max\left(\varepsilon_a\right)}\right)}{n_\chi\left(\chi(\mathbf{x}_\mathrm{max\left(\varepsilon_a\right)})\right)}=\frac{\sigma^\prime_f}{E}\cdot\left(2N_{i_\mathrm{det}}\left(\mathbf{x}_\mathrm{max\left(\varepsilon_a\right)}\right)\right)^b+\varepsilon^\prime_f\cdot\left(2N_{i_\mathrm{det}}\left(\mathbf{x}_\mathrm{max\left(\varepsilon_a\right)}\right)\right)^c\label{eq:Main_CMBNotchDet}
\end{align}
and the probabilistic median-W\"ohler curve predictions for the cooling hole specimen data sets at \unit{871}{\celsius}.

%\begin{figure}[htbp!]
	%\centering
%\input{Images/Main/IN939/M5003_cleanWoehlerPlot_StandKt2Kt3CH45CHALT_20170411_1018.tex}
%\caption[]{Prob. W\"ohler curves for \textit{IN-939}""(CC) specimens at \unit{649}{\celsius} and \unit{871}{\celsius}. Dashed lines represent fit curves of smooth and notched specimens. Solid lines represent prob. life predictions curves of cooling hole specimen ($K_t=4.0$ (\ref{fig:Main_CHALT_surfStress}) and $K_t=2.1$ (\ref{fig:Main_CH45_surfStress})). Dash-dotted lines represent deterministic life predictions of cooling hole specimens at \unit{871}{\celsius}.}
	%\label{fig:Main_M5003_clean1018_CH_Detcomp}
%\end{figure}

While \ref{eq:Main_CMBNotchDet} only evaluates life at the point of highest strain $\mathrm{max\left(\varepsilon_a\right)}$ the probabilistic curve predictions are calculated according to the surface integral \eqref{eq:The_ScaleSurfInt} and the notch support extension at every integration point as defined in \ref{eq:The_CMBwithNoS}.

\begin{figure}[htbp]
\begin{minipage}[t]{0.5\columnwidth}
	\centering
	\includegraphics[width=\textwidth]{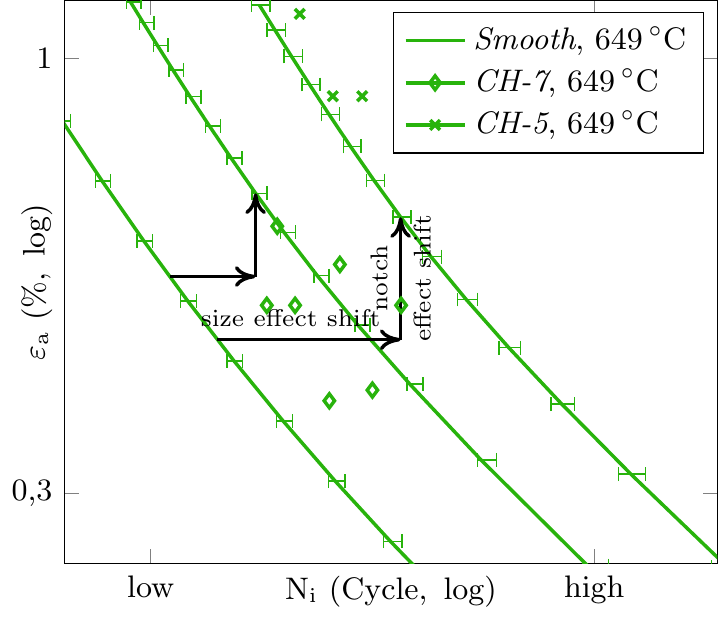}
	\end{minipage}
\begin{minipage}[t]{0.5\columnwidth}
	\centering
	\includegraphics[width=\textwidth]{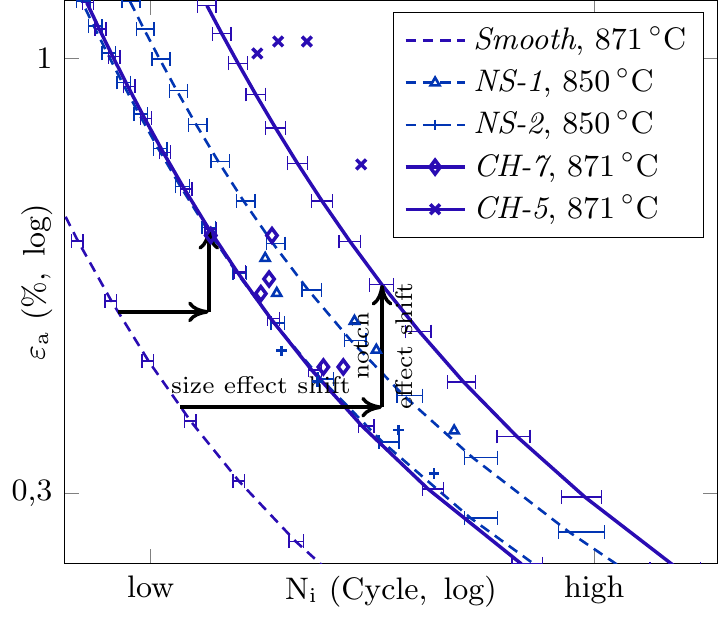}
	\end{minipage}
		\caption[]{Prob. W\"ohler curves and LCF test data points of \textit{IN-939} at \unit{649}{\celsius} (left) and \unit{871}{\celsius} and \unit{850}{\celsius} (right). The horizontal bars represent the \unit{92.5}{\%} confidence intervals of the curve prediction.}
	\label{fig:Main_IN939_Boot0411}
\end{figure}

\autoref{fig:Main_M5003_clean1018_CH_Detcomp} shows that deterministic LCF life predictions of the cooling hole specimens are over-conservative while the probabilistic \unit{871}{\celsius}-W\"ohler curves are shifted to significantly higher life in \autoref{fig:Main_M5003_clean1018_CH_Detcomp}. The predicted LCF life of the \textit{CH-7} specimen is shifted by a factor of five in total and by a factor 15 for the \textit{CH-5} specimen respectively. This is also true for the predictions at \unit{649}{\celsius} as shown with the green curves. These results point out that the local probabilistic approach considering the combined size and notch effect predicts LCF life much more accurately compared to the deterministic results for cooling hole geometries. Note that the ratio $n_\chi(\mathit{CH\text{-}5})/n_\chi(\mathit{CH\text{-}7})\approx 1.13$ is small and therefore does not explain the big difference in the predicted curve shifts and the test data position alone. The size effects factors however, $SE(\mathit{CH\text{-}5})=294$ and $SE(\mathit{CH\text{-}7})=24$, differ more than an order of magnitude and therefore explain why the \textit{CH-5} W\"ohler curve is shifted to much higher life than the \textit{CH-7} W\"ohler curve. The subdivision of both curve shifts is shown in \autoref{fig:Main_IN939_Boot0411}.

%\begin{figure}[htbp!]
	%\centering
%\input{Images/Main/IN939/M5003_CurveShifts649C.tex}
%\input{Images/Main/IN939/M5003_CurveShifts871C.tex}
%\caption[]{Fitted smooth specimen W\"ohler curve and predicted curves for both cooling hole specimen geometries (\textit{CH-5} and \textit{CH-7} specimen) at \unit{871}{\celsius}. The curve shift from the combined size and notch effect is subdivided into its contributions. The size effect shifts horizontally while the notch effect shifts vertically. The difference in the shifts from the notch support effect is far less significant than the difference in size effect.}
	%\label{fig:Main_M5003_shiftSubDiv}
%\end{figure}

%The predictions at high stress gradients are more conservative with the currently used set of CMB and notch support parameters.
%\begin{figure}[htbp!]
	%\centering
	%\input{Images/Main/IN939/M5003_BootPlot_StandKt2Kt3CH45CHALT649C_20170411_1018.tex}
	%\caption[]{Prob. W\"ohler curves and LCF test data points of \textit{IN-939} at \unit{649}{\celsius}. The horizontal bars represent the \unit{92.5}{\%} confidence intervals of the curve prediction.}
	%\label{fig:Main_IN939_Boot0411_649C}
%\end{figure}
%
%\begin{figure}[htbp!]
	%\centering
	%\input{Images/Main/IN939/M5003_BootPlot_StandKt2Kt3CH45CHALT871C_20170411_1018.tex}
	%\caption[]{Prob. W\"ohler curves and LCF test data points of \textit{IN-939} at \unit{871}{\celsius} and \unit{850}{\celsius}. The horizontal bars represent the \unit{92.5}{\%} confidence intervals of the curve prediction.}
	%\label{fig:Main_IN939_Boot0411_871C}
%\end{figure}

The significant difference of the size effect shifts does not only originate from the different number of cooling holes (five in the \textit{CH-5} geometry; seven in the \textit{CH-7} geometry) but even more from the cooling hole design. The image in the left box of \autoref{fig:Main_CH_surfStress} clearly shows that cooling holes perpendicular to the load axis (vertical) lead to far smaller areas of stress concentration compared to a geometry where they are in a \unit{45}{\degree} angle as it is shown in the right box of \autoref{fig:Main_CH_surfStress}. Ultimately, this demonstrates the importance of evaluating the combined size- and notch support effect in LCF and also the capability of the extended local probabilistic model to actually predict the combined effects.

In order to receive an estimate for the uncertainties of the curve predictions, parametric bootstrapping was conducted with the minimal number of 2000 bootstrap resamplings which is the minimum number concerning the statistical validity (see sections 9.2.1 and 9.2.2 of \cite{Escobar_Meeker1998}) and the maximum number concerning the computational effort for a 12-core desktop workstation. The shown confidence intervals in \autoref{fig:Main_IN939_Boot0411} are calculated with the percentile method as proposed in section 9.4.2 of \cite{Escobar_Meeker1998} and \cite{Efron1993} since the standard error computation for every bootstrap estimate is not feasible. Hence they represent the confidence in the median curve prediction and are not to mistaken with the residual scatter predicted by the estimated crack initiation life distribution. Calculating the confidence intervals with the percentile method is particularly expensive for the bootstrap curves of the cooling hole specimens, since one curve calculation with 14 curve points (14 surface integrals) takes $\approx\unit{2}{\min}$ for these more sophisticated geometries.

\subsection{\textit{Rene80} (HIP)}\label{subsec:Main_Rene80}
The second validation case of this work was built considering smooth specimen and notch specimen \textit{Rene80} (HIP) data at \unit{850}{\celsius}. It is shown that the extended probabilistic model for LCF can be calibrated with smooth specimen data points and data of only one notch specimen geometry. \textit{Rene80} (HIP) is a polycrystalline \ce{Ni}-based superalloy which is, like \textit{IN-939}, also used as gas turbine blade and vane base material where the combined notch support and size effect is influencing the LCF life \cite{ASME2017Paper}. 

\begin{table}[htbp!]
\centering% NICHT \begin{center}
\begin{tabular}{|c|c|c|c|c|c|c|c|c|c|c|}%\label{table.1.1}
  \hline
  Element & Ni & Cr & Co & Ti & Mo & W & Al & C & B & Zr \\
  \hline
  Weight - $\%$ & bal. & 14.0 & 9.5 & 5.0 & 4.0 & 4.0 & 3.0 & 0.17 & 0.15 & 0.03 \\
  \hline
\end{tabular}
\caption{Chemical composition of \textit{Rene80}. The examined specimens were cast and heat treated such that the material had a yield strength $YS_{0.2}\geq\unit{733}{\mega\pascal}$ at \unit{400}{\celsius}.}\label{tab:ChemCompRene80}
\end{table}
The average chemical composition of the evaluated \textit{Rene80} (HIP) specimens is given in \autoref{tab:ChemCompRene80}.

LCF data of four different specimen geometries was examined. The number of available test points for the respective geometry is listed in \autoref{tab:Main_Rene80_TestCollection}
%\vspace{5mm}
\begin{table}[htbp!]
\centering% NICHT \begin{center}
\begin{tabular}{|p{0.72\textwidth}|p{0.2\textwidth}|}
  \hline
	Geometry & $T_\text{Test}$\\
	\hline
	\hline
	Smooth cylindrical, $r_\text{Gauge}=\unit{5}{\milli\meter}$, \textit{Smooth-1} & \unit{850}{\celsius}\\
	\hline
  Small smooth cylindrical, $r_\text{Gauge}=\unit{3.5}{\milli\meter}$, \textit{Smooth-2}  & \unit{850}{\celsius}\\
	\hline
	Notched with $r_\text{Notch}=\unit{2.4}{\milli\meter}$, $K_t=1.62$, \textit{NS-1} & \unit{850}{\celsius}\\
	\hline
	Notched with $r_\text{Notch}=\unit{0.6}{\milli\meter}$, $K_t=2.60$, \textit{NS-2} & \unit{850}{\celsius}\\
	\hline
\end{tabular}
\caption{Available LCF test points for \textit{Rene80}.}\label{tab:Main_Rene80_TestCollection}
\end{table}
The geometry of the tested \textit{Rene80} notched specimens is sketched in \autoref{fig:Main_IN939_NotchSketch}. These and the small smooth specimens are from the same batch while the larger smooth specimens were manufactured in another batch\footnote{Siemens AG proprietary test data} but with equal heat treatment \cite{Beck_Gottschalk2014}. That is why only the small smooth specimen data set was used for calibration together with notch specimen points and the large smooth specimen data set was only used to compare to probabilistic life predictions. All tests were performed at \unit{850}{\celsius} such that no temperature model was required to interpolate the CMB parameters.

\begin{figure}[htbp]
\begin{minipage}[htb]{0.5\columnwidth}
	\centering
	\includegraphics[width=\textwidth]{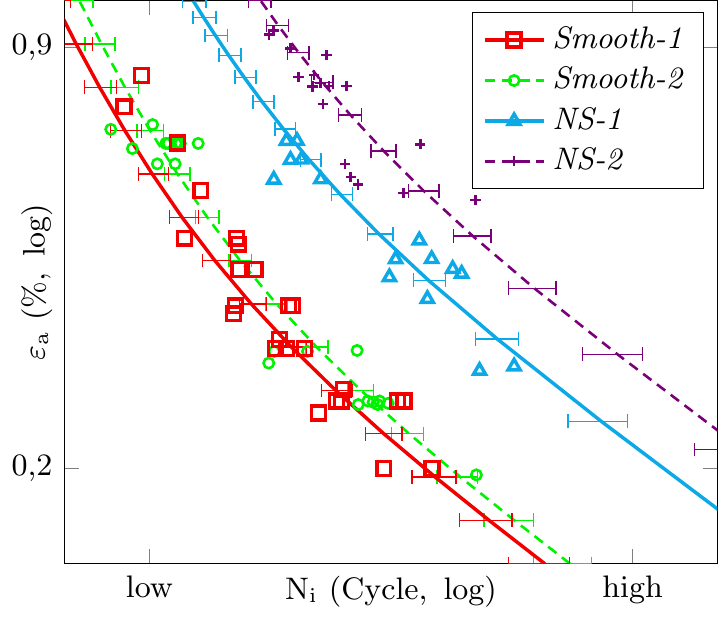}
	\end{minipage}
\begin{minipage}[htb]{0.5\columnwidth}
	\centering
	\includegraphics[width=\textwidth]{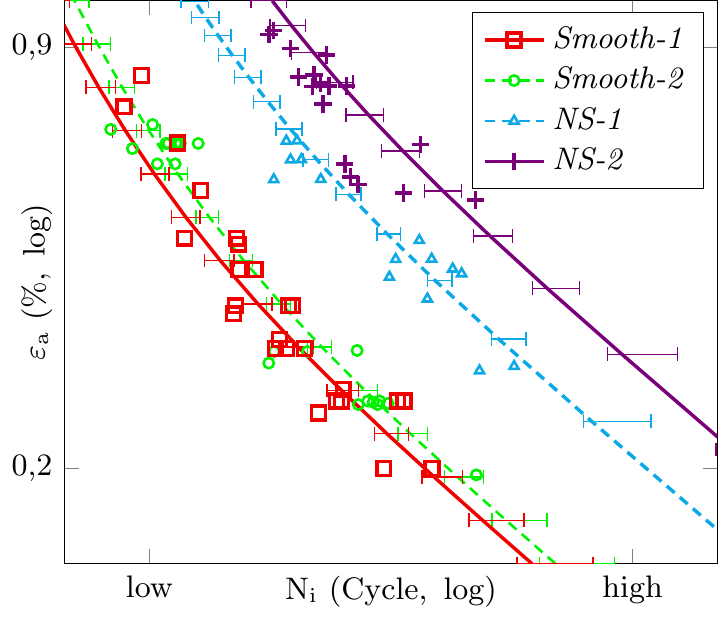}
	\end{minipage}
		\caption[]{Prob. W\"ohler curves and LCF test data points of \textit{Rene80} (HIP). The left plot shows the result from using test data points of notch specimen geometry \textit{NS-2} for fitting. The right plot shows the result from using test data points of notch specimen geometry \textit{NS-1}.}
\label{fig:Main_M0386_clean}
\end{figure}

%\begin{figure}[htbp!]
	%\centering
	%\input{Images/Main/Rene80/M0386_cleanWoehlerPlot_GStandardGsmStandGNotchGsmNot_20170407_0638.tex}	
	%\caption[]{Prob. W\"ohler curves and LCF test data points of \textit{Rene80}""(HIP). Test data points of sample geometry with conc. factor $K_t=2.6$ is used for fitting. The life prediction for sample geometry with conc. factor $K_t=1.62$ interpolates the notch support effect.}
	%\label{fig:Main_M0386_clean0638}
%\end{figure}
\autoref{fig:Main_M0386_clean} shows the fitted (thin, dashed) and predicted (thick, solid) probabilistic W\"ohler curves for LCF crack initiation life of smooth and notch specimen data. The predicted LCF life for the smooth specimen (red curves) is lower than the life of the small standard specimen (green) which firstly fits the data point distribution and secondly agrees with the theory of the statistical size effect, since the smooth specimens have a larger critical area. The curve prediction for the \textit{NS-2} specimen type in the right plot of \autoref{fig:Main_M0386_clean} is slightly too optimistic for the test points. Here, the notch effect is overestimated ($n_\chi=2.70$) compared to the curve prediction in the left plot ($n_\chi=2.43$). There, the Weibull shape $m$ is also smaller which leads to a higher size effect shift and hence a correct prediction for both notch specimen types.

%\begin{figure}[htbp!]
	%\centering
%\input{Images/Main/Rene80/M0386_cleanWoehlerPlot_GStandardGsmStandGNotchGsmNot_20170407_0728.tex}
%\caption[]{Prob. W\"ohler curves and LCF test data points of \textit{Rene80}""(HIP). Test data points of sample geometry with conc. factor $K_t=1.62$ is used for fitting. The life prediction for sample geometry with conc. factor $K_t=2.6$ extrapolates the notch support effect.}
	%\label{fig:Main_M0386_clean0728}
%\end{figure}

Comparing the negative Log-Likelihood sum for all test points confirms that the calibration which uses data of the \textit{NS-2} specimens yields to a better calibrated model.

\begin{table}[htbp!]
\centering% NICHT \begin{center}
\begin{tabular}{|c|c|}%\label{table.1.1}
  \hline
	Notch geometry used for calibration & $L=-\mathlarger{\sum_i}{\,\mathrm{log}\left(\frac{m}{\eta_i}\cdot \left(\frac{m}{\eta_i}\right)^{m-1}\cdot e^{-\left(\frac{n_i}{\eta_i}\right)^m}\right)}$ \\
	\hline
	\hline
	\textit{NS-2}, $r=\unit{0.6}{\milli\meter}$, $K_t=2.60$ & $L=769$ \\
  \hline
  \textit{NS-1}, $r=\unit{2.4}{\milli\meter}$, $K_t=1.62$ & $L=772$ \\
  \hline
\end{tabular}
\caption{Negative Log-Likelihood sum as measure for the goodness of fit compared for both calibrations conducted with \textit{Rene80} data.}\label{tab:Main_Rene80_ResultComp}
\end{table}
%Besides the aforementioned sensitivity of the Nelder-Mead-optimization to starting values (initial parameter set)
One cause for this difference might be the different scatter pattern in both notch data sets. The data point distribution from the \textit{NS-2} specimen testing seems to be more representative for the material batch and hence produces a more suitable parameter set when used for model calibration. 

%Regardless of these peculiarities
Nevertheless, both calibrations lead to models with good LCF life prediction reliability at \unit{850}{\celsius}, as the prediction curves lie almost in the middle of the scatter range of their respective data point cloud. From these results one can conclude that the simultaneous estimation of CMB and notch support parameters for calibrating the probabilistic model requires only LCF test data of one smooth and one notch specimen type for \textit{Rene80} (HIP).

The uncertainty in the curve prediction is again, equally to the \textit{IN-939} case, determined by 2000 bootstrap fits and all confidence intervals are calculated with the percentile method from those.

%\begin{figure}[htbp!]
	%\centering
	%\input{Images/Main/Rene80/M0386_BootPlot_GStandardGsmStandGNotchGsmNot_20170407_0638.tex}
	%\caption[]{Prob. W\"ohler curves and LCF test data points of \textit{Rene80}. The thin grey lines are Bootstrap curves (every 10th of the 2000) for the respective data points.}
	%\label{fig:Main_Rene80_Boot0638}
%\end{figure}
The deviation of the bootstrap curves from the master curve is larger in the Basquin branches (elastic deformation) than for the Coffin-Manson branches (plastic deformation) of the W\"ohler curves. The reason for that is the small amount of test points in that region. 
%A parameter set producing a curve with large deviation from the master curve in a certain strain band does not have a significant impact on the negative log-likelihood sum if only a small number of test points is available there. The consequence that prediction confidence intervals are broad for strain ranges with few test points is therefore very intuitive.
% Entfernt auf Empfehlung von Prof. Gottschalk
An equivalent situation but with a scarce data basis in the region of plastic deformation is observable in the \textit{26NiCrMoV 14-5} calibration case in the following \autoref{subsec:Main_Steel14-5}.

\subsection{\textit{26NiCrMoV 14-5}}\label{subsec:Main_Steel14-5}
A notch support effect is not only observed for superalloy specimens but also for \ce{Fe}-based steels \cite{Harders_Roesler2007}. The broad application range of steels in all industry sectors naturally crates an incentive for examining the combined size and notch effect for this material class. Here, data of the martensitic steel \textit{26NiCrMoV 14-5} is used to calibrate the local probabilistic model for LCF with respect to combined size and notch support effect. This steel is used for forged turbomachinery rotors discs which also have inhomogeneous stress concentrations at the steeple. Operating temperatures however are much lower compared to turbine blades and vanes. This is why test points collected at room temperature are analyzed for the first instance. The chemical composition of \textit{26NiCrMoV 14-5} is given in \autoref{tab:ChemCompM0044}.
%\vspace{5mm}
\begin{table}[htbp!]
\centering% NICHT \begin{center}
\begin{tabular}{|c|c|c|c|c|c|c|c|c|c|c|}%\label{table.1.1}
  \hline
  Element & Fe & Ni & Cr & Mo & V & Mn & C & Si \\
  \hline
  Weight - $\%$ & min. 93.5 & 3.7 & 1.5 & 0.35 & 0.1 & 0.28 & 0.27 & 0.05\\
  \hline
\end{tabular}
\caption{Chemical composition of \textit{26NiCrMoV 14-5}. The examined specimens were cast and heat treated such that the material had a yield strength $YS_{0.2}\geq\unit{510}{\mega\pascal}$ at \unit{400}{\celsius}.}\label{tab:ChemCompM0044}
\end{table}

Notch specimen LCF test points for seven different notch geometries leading stress concentration factors ranging from $K_t=1.28$ ($r_\text{Notch}=\unit{7}{\milli\meter}$) to $K_t=2.33$ ($r_\text{Notch}=\unit{1}{\milli\meter}$) have been available for examination. 
%\clearpage
The number of available test points for the respective geometry is listed in \autoref{tab:Main_M0044_TestCollection}.
%\vspace{5mm}
\begin{table}[htbp!]
\centering% NICHT \begin{center}
\begin{tabular}{|p{0.65\textwidth}|p{0.135\textwidth}|p{0.135\textwidth}|}
  \hline
	Geometry & $T_\text{Test}$ & Points\\
  \hline \hline
  smooth cylindrical $r_\text{Gauge}=\unit{5}{\milli\meter}$, \textit{Smooth} & \unit{20}{\celsius} & 5\\
	\hline
	Notched with $r_\text{Notch}=\unit{7}{\milli\meter}$, $K_t=1.28$, \textit{NS-1}  & \unit{20}{\celsius} & 1\\
	\hline
	Notched with $r_\text{Notch}=\unit{6}{\milli\meter}$, $K_t=1.32$, \textit{NS-2}  & \unit{20}{\celsius} & 6\\
  \hline
	Notched with $r_\text{Notch}=\unit{5}{\milli\meter}$, $K_t=1.38$, \textit{NS-3} & \unit{20}{\celsius} & 1\\
	\hline
	Notched with $r_\text{Notch}=\unit{4}{\milli\meter}$, $K_t=1.47$, \textit{NS-4}  & \unit{20}{\celsius} & 1\\
	\hline
	Notched with $r_\text{Notch}=\unit{3}{\milli\meter}$, $K_t=1.61$, \textit{NS-5}  & \unit{20}{\celsius} & 1\\
	\hline
	Notched with $r_\text{Notch}=\unit{2}{\milli\meter}$, $K_t=1.83$, \textit{NS-6}  & \unit{20}{\celsius} & 1\\
	\hline
	Notched with $r_\text{Notch}=\unit{1}{\milli\meter}$, $K_t=2.33$, \textit{NS-7} & \unit{20}{\celsius} & 20\\
	\hline
\end{tabular}
\caption{Available LCF test points for \textit{26NiCrMoV 14-5}.}
\label{tab:Main_M0044_TestCollection}
\end{table}
All specimens with notch radius $r_\text{Notch}>\unit{2}{\milli\meter}$ were of the geometry as shown in the upper sketch in \autoref{fig:Main_IN939_NotchSketch}. Specimens with $r_\text{Notch}\leq\unit{2}{\milli\meter}$, however, had a conical opening as transition from outside to the circular notch root instead of a parallel opening as shown in the lower sketch in \autoref{fig:Main_IN939_NotchSketch}. Again, all data points of different specimen geometries, could be used simultaneously for calibration and to validate life predictions since the extended local probabilistic model for LCF is entirely geometry independent. Only few smooth specimen data points exist. Most points have been collected with \textit{NS-2} specimens and \textit{NS-7} specimens. They show a remarkable scatter of almost half an order of magnitude. It is to presume that all other tested specimen types would exhibit a comparable scatter putting a significant uncertainty on every life prediction.
%Unfortunately only few smooth specimen data points exist for the specific strength class which resulted from the applied heat treatment of the \textit{26NiCrMoV""14-5} specimen cast.

\begin{figure}[htbp]
\begin{minipage}[htb]{0.5\columnwidth}
	\centering
	\includegraphics[width=\textwidth]{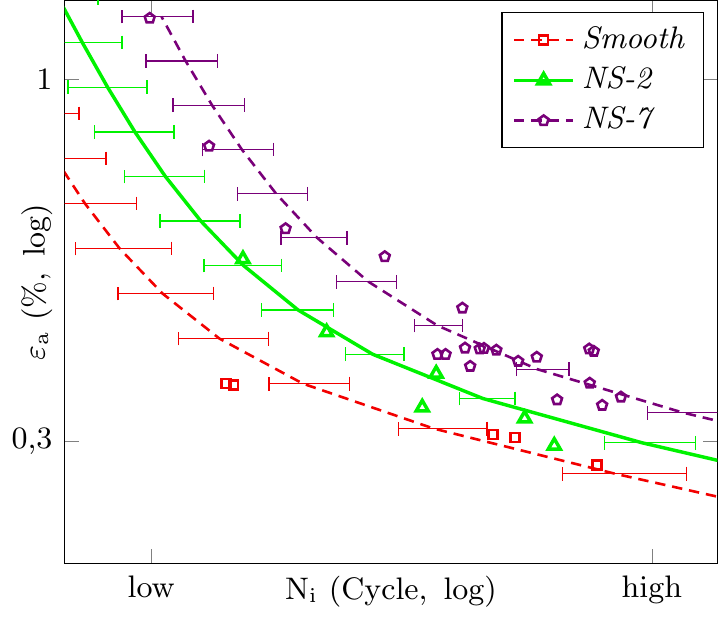}
	\end{minipage}
\begin{minipage}[htb]{0.5\columnwidth}
	\centering
	\includegraphics[width=\textwidth]{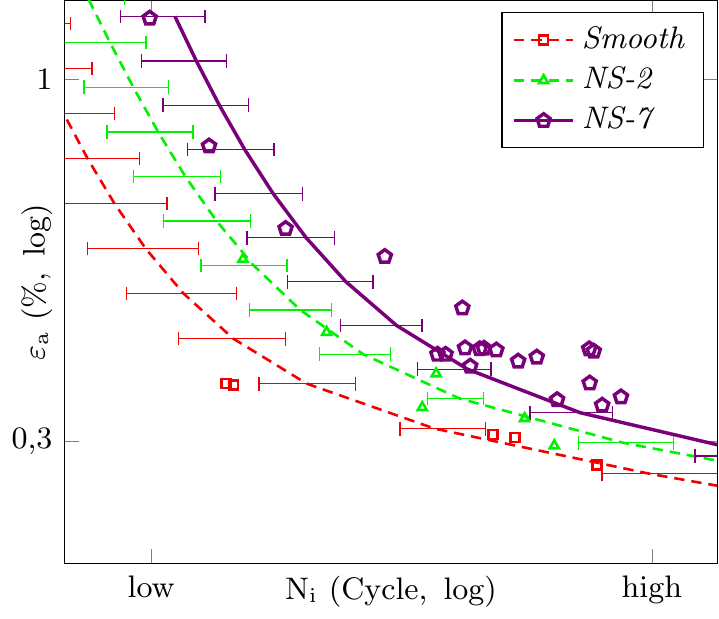}
	\end{minipage}
	\caption[]{Prob. W\"ohler curves and LCF test data points of \textit{26NiCrMoV 14-5}. The left plot
shows the result from using test data points of notch specimen geometry \textit{NS-7} for fitting. The right plot shows the result from using test data points of notch specimen geometry \textit{NS-2}.}
	\label{fig:Main_M0044_clean}
\end{figure}

Similar to the procedure in \ref{subsec:Main_Rene80}, two calibration cases were tested. The first calibration used the data points of the \textit{NS-7} specimens while the second used the \textit{NS-2} specimen data. Nevertheless, it is possible to also incorporate the single data points from \autoref{tab:Main_M0044_TestCollection} in the calibration procedure as they simply add another summand to the negative log-likelihood sum. Three of the single data points were added in each of the aforementioned calibrations but they are left out in \autoref{fig:Main_M0044_clean} where only the major fit and prediction curves are shown.

%\begin{figure}[htbp!]
	%\centering
	%\input{Images/Main/M0044/M0044_cleanWoehlerPlot_StandKt132Kt233_20170324_1329.tex}	
	%\caption[]{Prob. W\"ohler curves and LCF test data points of \textit{26NiCrMoV""14-5}. Test data points of sample geometry with conc. factor $K_t=2.33$ is used for fitting. The life prediction for sample geometry leading to $K_t=1.32$ interpolates the notch support effect.}
	%\label{fig:Main_M0044_clean1329}
%\end{figure}
%
%\begin{figure}[htbp!]
	%\centering
	%\input{Images/Main/M0044/M0044_cleanWoehlerPlot_StandKt132Kt233_20170329_1954.tex}	
	%\caption[]{Prob. W\"ohler curves and LCF test data points of \textit{26NiCrMoV""14-5}. Test data points of sample geometry with conc. factor $K_t=1.32$ is used for fitting. The life prediction for sample geometry leading to $K_t=2.33$ extrapolates the notch support effect.}
	%\label{fig:Main_M0044_clean1954}
%\end{figure}

As done for the previously assessed materials in \autoref{subsec:Main_IN939} and \autoref{subsec:Main_Rene80}, an uncertainty quantification for the probabilistic W\"ohler curve prediction is conducted. 
%Especially in the light of the high test point scatter this is of great importance here. Entfernt auf Empfehlung von Prof Gottschalk.
The confidence intervals become broader towards higher strains at the Coffin-Manson branches. As mentioned in \autoref{subsec:Main_Rene80}, the reason for that is the scarce data situation for strains $\varepsilon_a > \unit{0.5}{\%}$ where the negative log-likelihood sum, whose minimum is the convergence criterion in MLE, does not change significantly. 

The right plot of \autoref{fig:Main_M0044_clean} reveals that calibration using the data set of the \textit{NS-2} specimens yields to a too conservative LCF life prediction for the \textit{NS-7} specimen geometry. Extrapolating the combined notch support and size effect with this parameter set does not provide accurate results, particularly for elastic deformation. If points of the \textit{NS-7} specimens are used for calibration however, interpolating the combined effect on the \textit{NS-2} specimen geometry is acceptable (left plot in \ref{fig:Main_M0044_clean}). This observation is qualified by the very different values of the negative log-likelihood sums as listed in \ref{tab:Main_M0044_ResultComp}.
%\vspace{5mm}
\begin{table}[htbp!]
\centering% NICHT \begin{center}
\begin{tabular}{|c|c|}%\label{table.1.1}
  \hline
	Notch geometry used for calibration & $L=-\mathlarger{\sum_i}{\,\mathrm{log}\left(\frac{m}{\eta_i}\cdot \left(\frac{m}{\eta_i}\right)^{m-1}\cdot e^{-\left(\frac{n_i}{\eta_i}\right)^m}\right)}$ \\
	\hline
	\hline
	\textit{NS-7}, $r_\text{Notch}=\unit{1}{\milli\meter}$, $K_t=2.33$ & $L=287$ \\
  \hline
  \textit{NS-2}, $r_\text{Notch}=\unit{6}{\milli\meter}$, $K_t=1.32$ & $L=345$ \\
  \hline
\end{tabular}
\caption{Negative Log-Likelihood sum as measure for the goodness of fit compared for both calibrations conducted with \textit{26NiCrMoV 14-5} data. A deeper minimum is found when calibrating with the data points of the \textit{NS-7} specimens.}\label{tab:Main_M0044_ResultComp}
\end{table}

%However, the predicted W\"ohler curve describes the test points of for the \textit{NS-2} specimen geometry slightly worse (elastic regime) than in the corresponding fit curve in the right plot of \autoref{fig:Main_M0044_clean}. Apparently, the larger number of test points from \textit{NS-7} specimen tests dominates the curve trend to higher LCF life in the lower, elastic tail. Given the large scatter observed for both specimen types, it is expected that further testing of specimens with the \textit{NS-2} geometry would result in a similar data distribution. 
% Entfernt, da diese Beobachtung sehr schwach ist 

\begin{figure}[htbp]
%\begin{minipage}[htb]{0.5\columnwidth}
	\centering
	\includegraphics[width=0.5\textwidth]{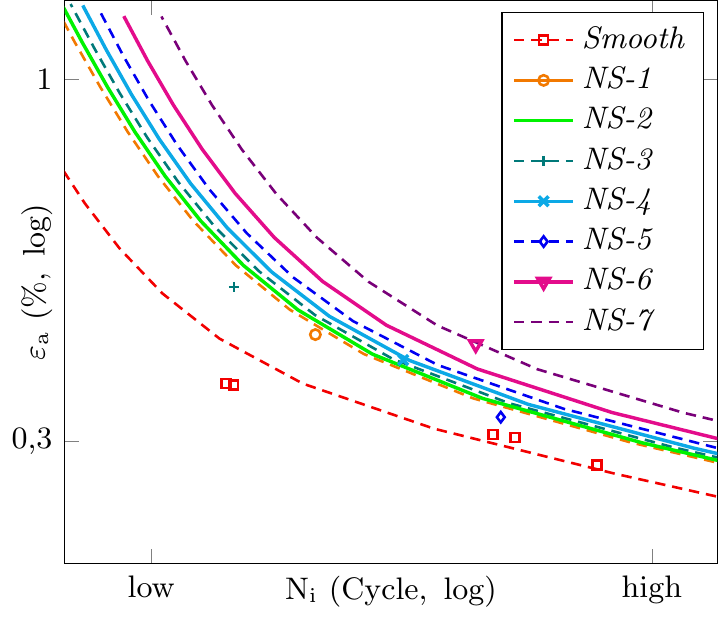}
%\end{minipage}
\caption[]{Prob. W\"ohler curves and LCF test data points for all tested geometries. The single test data points of the smooth specimens and specimen geometries \textit{NS-3}, \textit{NS-5} and \textit{NS-7} are used for fitting (dashed lines). Prob. W\"ohler curves for the other geometries (\textit{NS-1}, \textit{NS-2}, \textit{NS-4}, \textit{NS-6}) are predicted and drawn as solid lines. Test points of the \textit{NS-2} and \textit{NS-7} specimens are not shown.}
	\label{fig:Main_M0044_clean_all1329}
\end{figure}

Apparently, the model calibrated with \textit{NS-7} specimen data (sharply notched) is most accurate. This coincides with the observation in \autoref{subsec:Main_Rene80} where calibrating the model with sharply notched specimen data also lead to better predictions. \autoref{fig:Main_M0044_clean_all1329} shows W\"ohler curve predictions for the further tested single data points from different notch geometries. A certain mismatch between the test points of \textit{NS-4}, \textit{NS-5}, \textit{NS-6} specimens and their respective prediction curves is visible in \autoref{fig:Main_M0044_clean_all1329}. The reason for that is not necessarily a shortcoming of the model but very likely the test point scatter. 

As mentioned before, the test points of the \textit{NS-2} and \textit{NS-7} specimens are scattered across a broad life range and the crack initiation life of the other specimens from the same casting batch is expected to behave similarly. Aside from this, the comparison of curves to single points is potentially misleading but still shown for the sake of completeness here.

\subsection{Comparison of Notch Support mechanism in different materials} %\label{subsec:}
\begin{figure}[htbp!]
%\begin{minipage}[htb]{0.5\columnwidth}
	\centering
	\includegraphics[width=0.5\textwidth]{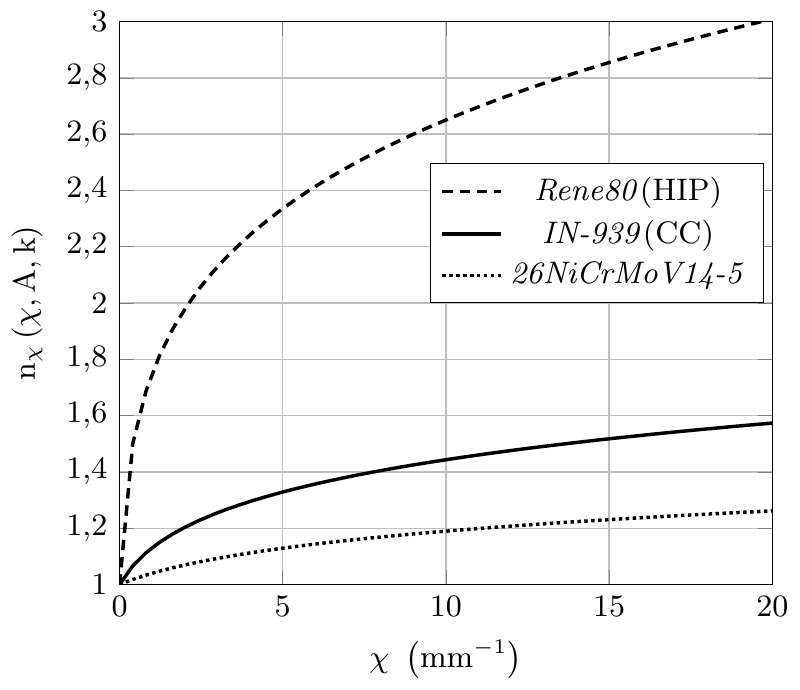}
%\end{minipage}
\caption[]{Plot of the notch support number $n_\chi$ vs. the normalized stress gradient $\chi$ for the tested materials.}
	\label{fig:Main_nChiComparison}
\end{figure}

Although the observed curve shift in the LCF W\"ohler curves always originates from the combined notch and size effect, the contributions can be clearly distinguished. \autoref{fig:Main_nChiComparison} shows plots of the notch support factor $n_\chi$ versus the normalized stress gradient $\chi$ for the three examined materials.

%\begin{figure}[htbp!]
	%\centering
	%\input{Images/Main/Chi_Comparison.tex}
	%\caption[]{Plot of the notch support number $n_\chi$ vs. the normalized stress gradient $\chi$ for the tested materials.}
	%%The two superalloys behave very different in the notch support effect. The life gain for \textit{Rene80}""(HIP) specimens is significantly higher than for \textit{IN-939}""(CC) specimens and the notch support for \textit{26NiCrMoV""14-5} is the lowest of all analyzed materials. These findings support the hypothesis that the notch support effect decreases with increasing ductility of the material.
	%\label{fig:Main_nChiComparison}
%\end{figure}

As visible in \autoref{fig:Main_nChiComparison}, the notch support factor is largest for \textit{Rene80} (HIP) and lowest for \textit{26NiCrMoV 14-5}. This coincides with the theory that the notch support effect is proportional to the materials grain size \cite{Harders_Roesler2007, Karry1953}. The reason for that are stresses dropping across the diameter of a critical surface grain retard the formation of persistent slip bands and hence macroscopic crack initiation. Smaller surface grains are always subject to higher lattice stress and therefore initiate cracks of a critical length after fewer load cycles. The large gap between $n_\chi$ for \textit{Rene80} (HIP) and \textit{IN-939} (CC) however cannot be explained with significant differences in the grain size. While Karry and Dolan provide a formula for the dependency of the fatigue notch factor $k_f$ on grain size in brass specimens \cite{Karry1953}, a robust relation between \textit{grain size} and the notch support parameters $A$ and $k$ for the geometry independent notch support factor $n_\chi$ is yet to be found. Note that $A$ and $k$ are single values and not explicitly considered to be statistically distributed as $k_f$ in \cite{Owolabi2015} for example. Instead, all additional data scatter emanating from the notch support effect is covered by the Weibull shape parameter $m$ during the model calibration.

%\begin{align}
%\begin{split}
%\end{split}
%\end{align}

%\section{}\label{sec:}
%
%%%%%%%%%%%%%%%%%%%%%%%%%%%%%%%%%%%%%%%%%%%%%%%%%%%%%%%%%%%%%%%%%%%%%%%
%\subsection{}\label{subsec:}
%
%\begin{figure}[htbp]
  %\centering
    %%\includegraphics[width=11.75cm]{.eps}
  %\caption{}
  %\label{fig:}
%\end{figure}

%%%%%%%%%%%%%%%%%%%%%%%%%%%%%%%%%%%%%%%%%%%%%%%%%%%%%%%%%%%%%%%%%%%%%%
%\section{}\label{sec:}

%% file: Discussion.tex
\section{Summarizing Discussion}
An extension for the probabilistic model for LCF, first presented in \cite{Schmitz_Seibel2013} and \cite{ASME2013Paper}, is described in \autoref{subsec:The_LCF_NSE}. It is now able to predict the LCF crack initiation life distribution of arbitrarily shaped parts and considers the combined size and notch support effect. This was first shown in \cite{ASME2017Paper} and is validated for three different materials (two \ce{Ni}-based superalloys, steel) in \autoref{sec:Main}. In every case, the model is simultaneously calibrated with LCF test data of standard and notched specimens since predicting the combined size and notch support effect requires all parameters (CMB, notch support and Weibull shape) to be adjusted to each other. In a second step, the calibrated model is validated by predicting the probabilistic LCF median W\"ohler curve for specimens with different geometries.

In the first case (\autoref{subsec:Main_IN939}), the extended model is calibrated with test data of \textit{IN-939}""(CC) specimens for different temperatures. Hence, temperature model functions for the CMB parameters are calibrated in a preliminary step. With those, the extended model is calibrated with respect to the combined size and notch effect using the maximum likelihood method. During the second step, the model is calibrated using LCF test data from smooth specimens and two different notch specimen types. In order to calculate the probabilistic life with the local approach from \cite{Schmitz_Seibel2013}, FEA model results of the notched specimens are used in every iteration. Even though this is computationally more costly, it creates the capability to calibrate the model with a reduced amount of test data (less different notch geometries). The calibrated model is able to predict the crack initiation lives of two different cooling hole specimens much more accurate than a deterministic approach which uses a CMB equation with notch support extension and neglects the statistical size effect. \autoref{fig:Main_IN939_Boot0411} shows that it is crucial to consider the combined size and notch support effect for reliable predictions. 

By predicting the cooling hole specimen lives, it is also shown how robust the extended local probabilistic model works concerning arbitrary geometries which is enabled by the local approach. Since the model is implemented in a highly automated way and does not require low level input information apart from the FEA model of the assessed component and material parameters, it is very suitable for structural optimization with respect to mechanical integrity. The primarily intended application however is robust component design with respect to scatter in LCF crack initiation life emanating from the uncertainties of material properties. As mentioned in \autoref{sec:Int}, Zhu \textit{et al.} have also approached the problem of probabilistic fatigue life prediction for multiaxially loaded parts \cite{ZhuBeretta2017}. In contrast to the local approach using the hazard density presented here, they performed latin hypercube sampling (LHS) of nonlinear 3D FEA calculations from probability distributions for the fatigue strength and ductility coefficient and used the resulting maximum shear strain and normal strain responses at the critical plane to calculate the fatigue life according to the Fatemi-Socie criterion \cite{FatemiSocie1988}. As an alternative to the LHS of FEA's, a first order approximation for the fatigue life distribution is provided as well. This would be more favorable for a possible application use of the method since the FEA computation times for complex geometries are usually too long for a LHS with $10^3$ samples. It would be very interesting to also validate the model at data which was not used for calibration, e.g. of cooling hole specimens in order to see the prediction capability of the Fatemi-Socie criterion for complex geometries.

In the second case (\autoref{subsec:Main_Rene80}), the extended model is calibrated with test data of \textit{Rene80}""(HIP) specimens tested at \unit{850}{\celsius}. Two model calibrations are conducted using smooth specimen data and data of two different notch specimen types, respectively. In both cases, the model is able to predict the probabilistic LCF life of the respective specimen type, whose data was not used for calibration. This shows that it is possible to use only tests of one notch specimen geometry, in addition to test points of smooth specimens to calibrate the extended model. A significantly reduced testing effort compared is feasible to the deterministic notch support parameter calibration. For the \textit{Rene80}""(HIP) case and in the third case (\textit{26NiCrMoV""14-5}) it is observed that models calibrated with test data from sharper notched specimen gives more accurate predictions. 

Furthermore, the confidence intervals of the curve predictions are computed from a set of bootstrap curves. Particularly for coarsely grained superalloys such as \textit{Rene80} the scatter of the test data is increased by grain orientation effects since it is very likely that only one single grain lies in the critical area in the notch root of the specimen. Its orientation is determining the actual value of the Young's modulus in load direction which generally deviates from the value assumed in the continuous mechanics approach in the local probabilistic model. Gottschalk \textit{et al.} introduced a strain amplitude correction scheme via the concept of Schmid factors for \textit{Rene80} data where the orientation of the failing grain was determined by \textit{electron backscatter diffraction} at every specimen \cite{Gottschalk_Schmitz2015}. This leads to a considerable reduction of data scatter. However, not only the local anisotropy of material properties but also the variation in granular strain due to the different composition of neighboring grains and their load transfer behavior requires a description beyond continuum mechanics \cite{Seibel_Diss}.
%The curve uncertainty does not exceed the residual scatter of the test points and therefore gives confidence in the calibration result. Entfernt auf Empfehlung von Prof. Gottschalk

Steels can also show a notch support effect where its intensity decreases with increased ductility and grain size. The available LCF test data of \textit{26NiCrMoV""14-5} at \unit{20}{\celsius} indicates a notch support effect and is used for calibration of the local probabilistic model. Analogously to the \textit{Rene80}""(HIP) case, two calibrations using data of two different notch specimen types are conducted and both model variants are able to predict an acceptable LCF W\"ohler curve of the notch specimen left out in calibration. Again, analogously to the \textit{Rene80}""(HIP) case, the model calibrated with the data set of the sharper notch geometry is predicting the LCF life of specimens which were not used for calibration more precisely. 

An alternative model could be implemented with a volume integration approach instead of the currently used surface integration. The stress gradient information would then be implicitly contained in the volume hazard density field.

One can conclude that the extended local probabilistic model for LCF could be more comprehensively validated due to its extension for taking into account the combined statistical size and notch effect. It is able to perform the fatigue assessment of more arbitrarily shaped parts at arbitrary load and considers material property variations by providing a distribution for the number of cycles to crack initiation.

An additional advantageous property of the model that has not been mentioned yet, is the robustness to stress deviations. The smoothness of the probabilistic LCF life functional with respect to meshing is particularly useful in structural optimization where fast, gradient based algorithms can be used.